\documentclass[english,11pt]{article}
\usepackage{style}
\begin{document}

\title{\textbf{On stability of power-law solution in multidimensional Gauss-Bonnet cosmology}}
\author{D.M.Chirkov\textsuperscript{1,2}\footnote{E-mail: \url{chirkovdm@live.com}}, A.V.Toporensky\textsuperscript{1}\footnote{E-mail: \url{atopor@rambler.ru}}}

\maketitle
\vspace{-.7cm}

\begin{center}
\textsuperscript{1}Sternberg Astronomical Institute, Moscow State University,\\ Universitetsky pr., 13, Moscow, 119991, Russia
\vskip .4cm

\textsuperscript{2}Faculty of Physics, Moscow State University,\\ Leninskie Gory, Moscow, 119991, Russia

\end{center}

\begin{abstract}
We consider dynamics of a flat anisotropic multidimensional cosmological model in Gauss-Bonnet gravity in the
presence of a homogeneous magnetic field. In particular, we find conditions under which the known power-law
vacuum solution can be an attractor for the case with non-zero magnetic field. We also describe a particular
class of numerical solution in $(5+1)$-dimensional  case which does not approach the power-law regime.
\end{abstract}

\section{Introduction.}
The fact that structure of initial cosmological singularity can be rather
complicated have been recognized since 60-th of the last century when the
conception of BKL chaos have been presented~\cite{BKL}. It appears that Kasner solution
being a general solution for a vacuum Bianchi I Universe becomes unstable
in the case of metric with positive spatial curvature (belonging to Bianchi IX class)
and is replaced by a complicated sequence of transient "Kasner epochs".
Later it was found that some classes of an anisotropic matter can induce similar
type of cosmological behavior even in flat Bianchi I case. This can be shown for
a magnetic field by LeBlanc~\cite{LeBlanc} and for a general vector field by Kirillov~\cite{Kirillov}.
The BKL analogs for the flat magnetic Universe is a typical behavior when off-diagonal terms in the metric
in the frame determined by magnetic field are present. For a diagonal case see for example \cite{Rosen}.

Possible generalisation of these results may involve studies in the framework
of multidimensional cosmology as well as alternative gravity theories.
Analogs of magnetic field for multidimensional space-times have been
studied in~\cite{Mitskievich1, Mitskievich2, Mitskievich3}. If we want to introduce some corrections to Einstein gravity, the first
goal is to find a solution which replaces Kasner solution for a flat anisotropic
Universe. In the present paper we consider Lovelock gravity as a generalization
of General Relativity. The principal feature of Lovelock gravity is that this
theory keeps the order of equations of motion the same as in GR, while other
theories (like popular now $f(R)$-theory) results in increasing of the number
of derivatives in equations of motion. Another important property of Lovelock gravity
is that it gives corrections to GR only in higher-dimensional space-time, so
it is natural to consider it in the framework of multidimensional cosmology.
As the number of non-Einstein terms in Lovelock gravity is finite for any
given dimensionality of space-time, it is possible to consider regimes when
these terms are not only  small corrections to Einstein gravity. In particular,
it is reasonable to expect that the highest order Lovelock term, consisting of
highest power of curvature invariants, dominates near a cosmological singularity.
In the present paper we consider the second Lovelock term, which is the famous
Gauss-Bonnet combination. It is the highest possible Lovelock term for $(4+1)$ and
$(5+1)$ dimensional spacetimes. As cosmology in the $(4+1)$ dimensional case has
some pathological features, we restrict ourselves by the $(N+1)$ dimensional case with $N\geqslant5$ in
the present paper.

In late 80-th some vacuum solutions for a flat anisotropic Universe in Gauss-Bonnet gravity for $N=4,5$ have been found~\cite{Deruelle1,Deruelle2}. They replace Kasner solution of GR. Later these solutions have been rediscovered and verified for $N=6,7$ in~\cite{Toporensky};  after that this solution have been generalized for all $N$ and also to the general Lovelock gravity~\cite{Pavluchenko}. The main goal of this paper is to address question of its stability near the initial singularity. It is known that in the presence of an ordinary isotropic matter these solutions are stable to the past if the matter has the equation of state with $w<1/3$, otherwise the solution tends to isotropic one~\cite{Kirnos1, Kirnos2}. For more interesting regimes near a singularity it is necessary to consider either anisotropic matter or curved geometries. As introduction of spatial curvature usually leads to very cumbersome equations of motion, we have chosen a magnetic fields as a possible source of instability.
In this paper we restrict ourselves by the diagonal case.

It is also known that in contrast to Einstein gravity certain initial conditions lead to exponential solutions instead to those of power-law behavior~\cite{Kirnos1,Ivashchuk1,Ivashchuk2}, we do not consider them in the present paper.

The structure of the paper is as follows: In Sec.~\ref{preliminaries} we describe the metric and matter content of the model studied, in Sec.~\ref{3}
the model is investigated in the framework of Einstein relativity. We present results on stability of Kasner solution in the
presence of magnetic field. The same model in Gauss-Bonnet gravity is studied in Sec.~\ref{4} with presentation made as parallel
as possible to presentation in the preceding section in order to compare Einstein and Gauss-Bonnet cases. In Sec.~\ref{5} we
describe  a particular regime existing in the zone of instability of power-law regimes in $(5+1)$-dimensional Gauss-Bonnet
gravity. Sec.~\ref{6} contains a brief summary of results obtained.

\section{Preliminaries.\label{preliminaries}}
In what follows we use a reference system chosen so that\footnote{Here and after Greek indices run from 0 to N and Latin indices from 1 to N.}: \eq{g_{00}=1,\quad g_{kk}=-e^{2a_k(t)},\quad g_{ij}=0,\quad i\ne j\label{g}}
In this section we looking for the energy-momentum tensor of a pure magnetic field in $(N+1)$-dimensional space-time. In the general case components of the energy-momentum tensor of the electromagnetic field are
\eq{T^{\mu}_{\nu}=\frac{1}{4\pi}\left(F_{\nu\gamma}F^{\gamma\mu}+\frac{1}{4}\delta^{\mu}_{\nu}F_{\alpha\beta}F^{\alpha\beta}\right)\label{T}}
where $F_{\alpha\beta}$ is the Faraday tensor. The components of the Faraday tensor obey the following equations:
\eq{\left\{\begin{array}{rcl}
\nabla_{\mu}F^{\mu\nu} & = & 0\,, \\
dF & = & 0 \,,
\end{array}\right.}
or, in more detail:
\eq{\left\{\begin{array}{l}
\eD_{\mu}F^{\mu\nu}+\Gamma^{\mu}_{\mu\lambda}F^{\lambda\nu}+\Gamma^{\nu}_{\mu\sigma}F^{\mu\sigma}=0\,, \\
(dF)_{\alpha\beta\gamma}=0,\quad \alpha<\beta<\gamma \,,\\
\end{array}\right.\label{sys}}
where $\eD_{\mu}=\frac{\partial}{\partial x^{\mu}}$. As it is known
\eq{(dF)_{\alpha_1\alpha_2\alpha_3}=\eD_{\alpha_1}F_{\alpha_2\alpha_3}-\eD_{\alpha_2}F_{\alpha_1\alpha_3}+\eD_{\alpha_3}F_{\alpha_1\alpha_2}\,,} therefore we may rewrite system~(\ref{sys}) in the form
\eq{\left\{\begin{array}{l}
\eD_{\mu}F^{\mu\nu}+\Gamma^{\mu}_{\mu\lambda}F^{\lambda\nu}+\Gamma^{\nu}_{\mu\sigma}F^{\mu\sigma}=0\,, \\
\eD_{\alpha}F_{\beta\gamma}-\eD_{\beta}F_{\alpha\gamma}+\eD_{\gamma}F_{\alpha\beta}=0,\quad \alpha<\beta<\gamma
\end{array}\right.\label{F}} The components of the Riemann connection have the form:
\eq{\Gamma^{i}_{i0}=\Gamma^{i}_{0i}=\dot{a}_i(t),\quad \Gamma^{0}_{jj}=\dot{a}_j(t)e^{2a_j(t)}\label{Riem-conn}}
Its other components are equal to zero. Since we consider the case of a homogenous space, Faraday tensor depends on time $t$ only; then, in view of~(\ref{Riem-conn}) the system~(\ref{F}) takes the form:
\eq{\left\{\begin{array}{l}
\dot{F}^{0i}+F^{0i}\sum\limits_{k}\dot{a}_k=0\,, \\
\dot{F}_{ij}=0 \end{array}\right.}
Its solutions are
\eq{\left\{\begin{array}{l}
F^{0m}=\phi_m\,e^{-\sum\limits_{k}a_k},\; \phi_m=\const\,, \\
F_{ij}=\psi_{ij},\; \psi_{ij}=\const \end{array}\right.\label{psi}}
Hereinafter we will be interested in the case of a pure magnetic field; so that $\phi_m=0$ and, as a consequence,
\eq{F^{0m}=F_{0m}=0\label{F0m}}
In what follows we will deal with diagonal energy-momentum tensor; in view of~(\ref{g}),(\ref{T}),(\ref{psi}) and (\ref{F0m}) it implies that
\eq{T^{\mu}_{\nu}=0,\;\mu\ne\nu\quad\Longleftrightarrow\quad F_{\nu\gamma}F^{\gamma\mu}=0,\;\mu\ne\nu\quad\Longleftrightarrow\quad\sum\limits_{l}e^{-2a_l}\psi_{il}\psi_{lk}=0\label{nondiag-T=0}}
Functions $e^{-2a_l}$ are linearly independent, therefore from~(\ref{nondiag-T=0}) it follows that
\eq{\psi_{il}\psi_{lk}=0,\quad i,k,l=\overline{1,N},\; i \ne k\label{psipsi}}
Each $\psi_{ij}$ is multiplied by all that $\psi_{kl}$, which has one of the indices $k,l$ coincident with one of the indices $i,j$.

Let us fix pair $(i,j)$; the number of combinations, in which $\psi_{ij}$ is found, equals to $2(N-2)$. Let $\psi_{ij}\ne 0$; then the other $2(N-2)$ quantities $\psi_{ik}$\, ($\psi_{jk}$), which are multiplied by $\psi_{ij}$, must be equals to zero; indeed, if $\psi_{ik}\ne 0\: (k\ne i,j)$, then $\psi_{ij}\psi_{ik}\ne 0$, but that contradict (\ref{psipsi}).

Continuing the argument, we find the number of zero magnetic field components:
\eq{2(N-2)+2(N-4)+\ldots =2\frac{(N-2)+(N-2\frac{N-1}{2})}{2}\,\frac{N-1}{2}=\frac{(N-1)^2}{2}\quad \mbox{for an odd dimensions}}
\eq{2(N-2)+2(N-4)+\ldots =2\frac{(N-2)+(N-2\frac{N}{2})}{2}\,\frac{N}{2}=\frac{N(N-2)}{2}\quad \mbox{for an even dimensions}}
We took into account that the number of pairs of indices without the same elements equals to $\frac{N-1}{2}$ for an odd dimensions and to $\frac{N}{2}$ for an even dimensions.

The total number of components of the magnetic field is $\frac{N(N-1)}{2}$; then the number $\chi$ of non-zero magnetic field components is
\eq{\chi=\frac{N(N-1)}{2}-\frac{(N-1)^2}{2}=\frac{N-1}{2}\quad \mbox{for an odd dimensions},\label{chi-odd}}
\eq{\chi=\frac{N(N-1)}{2}-\frac{N(N-2)}{2}=\frac{N}{2}\quad \mbox{for an even dimensions}\label{chi-even}}
In what follows they are exactly $\psi_{12},\psi_{34},\ldots,\psi_{\sss 2n-1,\,2n},\;n=\overline{1,\chi}$ that we set to be non-zero; other components of the Faraday tensor are assumed to be zero. Thus, components of the energy-momentum tensor of the electromagnetic field take the form:
\eq{T^{\mu}_{\nu}=0,\;\mu\ne\nu,\label{Tmunu}}
\eq{T^0_0=\sum\limits_{i=\overline{1,\chi}}\psi_{\scriptscriptstyle 2i-1,\,2i}^2e^{-2(a_{2i-1}+a_{2i})},\quad T^n_n=-\psi_{\sss 2n-1,\,2n}^2e^{-2(a_{2n-1}+a_{2n})}+\sum\limits_{j=\overline{1,\chi} \atop j\ne n}\psi_{\sss 2j-1,\,2j}^2e^{-2(a_{2j-1}+a_{2j})},\; n=\overline{1,\chi}\label{T00-Tnn}}
Hereinafter it will be convenient to use notations like this:
\eq{<n>=\bigl\{2n-1,2n\bigr\},\quad n=\overline{1,\chi}\label{<n>}}
We assume that
\eq{i\ne <n>\quad\Longleftrightarrow\quad i\ne 2n-1\;\wedge\;i\ne 2n,\qquad k=<n>\quad\Longleftrightarrow\quad k=2n-1\;\vee\;k=2n}
\section{Stability of the Kasner solutions.\label{3}}
\subsection{Field equations.}
The action reads:
\eq{S=\frac{1}{16\pi}\int d^{N+1}x\sqrt{|\det(g)|}\bigl(\mathcal{L}_{E}+\mathcal{L}_m\bigr)}
\eq{\mathcal{L}_{E}=R,\quad\mathcal{L}_m=F_{\alpha\beta}F^{\alpha\beta}}
where $R$ and $F_{\alpha\beta}$ are scalar curvature and Faraday tensor respectively. Here and after we use Planck units: \eq{c=1,\: G=\frac{1}{m_{Pl}^{N-2}}=1\,,}
$m_{Pl}$ is the Planck mass. The gravitational equations is given by
\eq{G^{\mu}_{\nu}=8\pi T^{\mu}_{\nu}\label{grav-eq}} where
\eq{G^{\mu}_{\nu}=R^{\mu}_{\nu}-\frac{1}{2}\delta^{\mu}_{\nu}R}
are the components of the Einstein tensor. Taking into account~(\ref{g}) we get:
\eq{G^{\mu}_{\nu}=0,\;\;\mu\ne\nu\label{Gmn}}
\eq{G^0_0=\sum\limits_{i<j}\dot{a}_i\dot{a}_j,\qquad G^n_n=\sum\limits_{i\ne n}\bigl(\ddot{a}_i+\dot{a}_i^2\bigr)+\sum\limits_{i<j \atop i,j\ne n}\dot{a}_i\dot{a}_j,\quad n\oneN\label{G00-Gnn}}
As we consider the diagonal case here, then in view of~(\ref{Tmunu}),(\ref{T00-Tnn}),(\ref{grav-eq}),(\ref{G00-Gnn}) for an even number of dimensions the gravitational equations can be written as
\eq{\sum\limits_{i\ne <n>}\bigl(\ddot{a}_i+\dot{a}_i^2\bigr)+\sum\limits_{i<j \atop i,j\ne <n>}\dot{a}_i\dot{a}_j=-\psi_{\sss 2n-1,\,2n}^2e^{-2(a_{2n-1}+a_{2n})}+\sum\limits_{j=\overline{1,\chi} \atop j\ne n}\psi_{\sss 2j-1,\,2j}^2e^{-2(a_{2j-1}+a_{2j})},\; n=\overline{1,\chi}\label{grav-eq-basic-even}}
\eq{\sum\limits_{i<j}\dot{a}_i\dot{a}_j-\sum\limits_{j=\overline{1,\chi}}\psi_{\sss 2j-1,\,2j}^2e^{-2(a_{2j-1}+a_{2j})}=0\label{first-int}}
Note that each expression like~(\ref{grav-eq-basic-even}) describes two equations with numbers $2n-1$ and $2n$ simultaneously (we use the notation~(\ref{<n>}) here). In the case of an odd number of dimensions there is one more equation in addition to these:
\eq{\sum\limits_{i\ne N}\bigl(\ddot{a}_i+\dot{a}_i^2\bigr)+\sum\limits_{i<j \atop i,j\ne N}\dot{a}_i\dot{a}_j=\sum\limits_{j=\overline{1,\chi}}\psi_{\sss 2j-1,\,2j}^2e^{-2(a_{2j-1}+a_{2j})},\; n=\overline{1,\chi}\label{grav-eq-basic-odd}}
The left side of the expression~(\ref{first-int}) is a first integral of~(\ref{grav-eq-basic-even}).
\subsection{Vacuum solution.}
When there is no any matter the equations~(\ref{grav-eq-basic-even})-(\ref{grav-eq-basic-odd}) lead to:
\eq{\sum\limits_{i\ne n}\bigl(\ddot{a}_i+\dot{a}_i^2\bigr)+\sum\limits_{i<j \atop i,j\ne n}\dot{a}_i\dot{a}_j=0,\quad n\oneN\label{grav-eq-basic-vac}}
\eq{\sum\limits_{i<j}\dot{a}_i\dot{a}_j=0\label{first-int-vac}}
These equations has the following solutions~\cite{Kasner}:
\eq{a_k=p_k\ln(t)+C_k,\quad C_k=\const,\; k\oneN,\qquad \sum\limits_{n}p_n^2=1,\quad\sum\limits_{n}p_n=1\label{Kasner-sol}}
As a consequence, components of the metric tensor follows a power law:
\eq{g_{kk}(t)=\e^{2a_k(t)}=\widetilde{C}_kt^{2p_k},\quad \widetilde{C}_k=\const,\quad k\oneN\label{Kasner-metr}}
This is well-known multidimensional Kasner solution.
\subsection{Stability conditions.\label{Einstein-stab.cond}}
Metric ceases to obey a power law when there is a matter; but it turns out that the metric of the space filled with the magnetic field may get close to the Kasner metric and converge to it when moving towards the initial singularity. Namely, numerical calculations shows that there exists solutions of the equations~(\ref{grav-eq-basic-even}) that get close to the Kasner solutions~(\ref{Kasner-sol}) and converge to them as time tends to the point $t=0$. Considering an influence of the magnetic field as a perturbation we will search for solutions of the equations~(\ref{grav-eq-basic-even}) in the form:
\eq{a_k(t)=a_k^0(t)+\varphi_k(t),\quad k\oneN\label{dist-Kasner}}
where $a^0_k$ is the Kasner solution~(\ref{Kasner-sol}), $\varphi_k \in C^2(\mathbb{R})$ is a perturbation. We will say that solutions $a_k^0$ are asymptotically stable if
\eq{\lim\limits_{t\rightarrow0}a_k(t)=a_k^0(t),\quad k\oneN}
In other words, solutions $a_k^0$ are asymptotically stable if and only if
\eq{\lim\limits_{t\rightarrow0}\varphi_k(t)=0,\quad k\oneN}
Now we will find out conditions which specify asymptotically stable Kasner solutions.

\noindent\textbf{Proposition 1.} \emph{Kasner solutions are asymptotically stable for $t\rightarrow0$ if and only if
$p_{2n-1}+p_{2n}<1,\;n=\overline{1,\chi}$.}\vspace{.2cm}
\pfi We consider the space of an even dimension; that of an odd dimension is treated similar way.\vspace{.1cm}

\noindent 1. Let $\{a_1^0,\ldots,a_N^0\}$ be an asymptotically stable solution of the equations~(\ref{grav-eq-basic-vac})-(\ref{first-int-vac}) given by~(\ref{Kasner-sol}); then there exists solution $\{a_1,\ldots,a_N\}$ of the equations~(\ref{grav-eq-basic-even}),(\ref{first-int}) and functions $\varphi_1,\ldots,\varphi_N\in C^2(\mathbb{R})$ such that
\eq{a_k(t)=a_k^0(t)+\varphi_k(t),\quad k\oneN\label{Kasner-disturbed}}
\eq{\lim\limits_{t\rightarrow0}\varphi_k(t)=0,\quad k\oneN\label{Kasner-phi}}
Substitution~(\ref{Kasner-sol}),(\ref{Kasner-disturbed}) to the equations~(\ref{grav-eq-basic-even}) and~(\ref{first-int}) leads to:
\eqs{\sum\limits_{\sss i\ne <n>}p_i(p_i-1)+\sum\limits_{\sss j<k \atop j,k\ne <n>}p_j p_k+t\left[\sum\limits_{\sss i\ne <n>}p_i\dot{\varphi}_i+\sum\limits_{\sss j<k \atop j,k\ne <n>}\Bigl(p_j\dot{\varphi}_k+p_k\dot{\varphi}_j\Bigr)\right]+t^2\left[\sum\limits_{\sss i\ne <n>}\Bigl(\ddot{\varphi}_i+\dot{\varphi}_i^2\Bigr)+\sum\limits_{\sss j<k \atop j,k\ne <n>}\dot{\varphi}_j \dot{\varphi}_k\right]=}
\eq{=-\psi_{\sss 2n-1,\,2n}^2e^{-2(\varphi_{2n-1}+\varphi_{2n})}t^{2-2(p_{2n-1}+p_{2n})}+
\sum\limits_{j=\overline{1,\chi} \atop j\ne n}\psi_{\sss 2j-1,\,2j}^2e^{-2(\varphi_{2j-1}+\varphi_{2j})}t^{2-2(p_{2j-1}+p_{2j})}\label{Kasner-trans-sys}}
\eq{\sum\limits_{i<j}p_i p_j+\sum\limits_{i<j}\Bigl(p_i\dot{\varphi}_j+p_j\dot{\varphi}_i\Bigr)+\sum\limits_{i<j}\dot{\varphi}_i\dot{\varphi}_j=
\sum\limits_{i=\overline{1,\chi}}\psi_{\scriptscriptstyle 2i-1,\,2i}^2e^{-2(\varphi_{2j-1}+\varphi_{2j})}t^{2-2(p_{2i-1}+p_{2i})}\label{Kasner-trans-constr}}
It follows from~(\ref{Kasner-phi}) that
\eq{\lim\limits_{t\rightarrow0}\bigl(t\dot{\varphi}_k(t)\bigr)=0,\quad\lim\limits_{t\rightarrow0}\bigl(t^2\ddot{\varphi}_k(t)\bigr)=0,\quad k\oneN\label{Kasner-phi-dot}}
Let $t\rightarrow0$; in view of~(\ref{Kasner-sol}),(\ref{Kasner-phi}),(\ref{Kasner-phi-dot}) equations~(\ref{Kasner-trans-sys})-(\ref{Kasner-trans-constr}) take the form:
\eq{0=-\psi_{\sss 2n-1,\,2n}^2t^{2-2(p_{2n-1}+p_{2n})}+\sum\limits_{j=\overline{1,\chi} \atop j\ne n}\psi_{\sss 2j-1,\,2j}^2t^{2-2(p_{2j-1}+p_{2j})}\label{Kasner-lim-sys}}
\eq{0=\sum\limits_{i=\overline{1,\chi}}\psi_{\scriptscriptstyle 2i-1,\,2i}^2t^{2-2(p_{2i-1}+p_{2i})}\label{Kasner-lim-constr}}
Equalities~(\ref{Kasner-lim-sys}) and~(\ref{Kasner-lim-constr}) are satisfied if and only if
\eq{p_{2n-1}+p_{2n}<1,\quad n=\overline{1,\chi}\label{ineq}}
q.e.d.

2. Let $\{a_1^0,\ldots,a_N^0\}$ be the Kasner solution given by~(\ref{Kasner-sol}) such that $p_{2n-1}+p_{2n}<1,\;n=\overline{1,\chi}$, and $\varphi_1(t),\ldots,\varphi_N(t)\in C^2(\mathbb{R})$ be a small deviations from that solution. We have:
\eq{a_k(t)=a_k^0(t)+\varphi_k(t),\quad k\oneN\label{Kasner-disturbed-1}}
Let $t=t_0>0$ be the initial moment; we assume that
\eq{|\varphi_i(t_0)|\ll |a_k^0(t_0)|,\quad |\dot{\varphi}_i(t_0)|\ll \left|\dot{a}_k^0(t_0)\right|,\quad i\oneN\label{Kasner-dev}}
Substitution~(\ref{Kasner-sol}),(\ref{Kasner-disturbed-1}) to the equations~(\ref{grav-eq-basic-even}) and~(\ref{first-int}) give us equations~(\ref{Kasner-trans-sys})-(\ref{Kasner-trans-constr}). In view of~(\ref{Kasner-dev}) we can neglect terms contained factors like $t\dot{\varphi}_i,\;t^2\dot{\varphi}_i\dot{\varphi}_j$ in the lhs of the equations~(\ref{Kasner-trans-sys})-(\ref{Kasner-trans-constr}) and terms contained $\varphi_i$ in the rhs of that equations. Namely, let $\varepsilon$ be a small positive real number; then in the $\varepsilon$-vicinity of the point $t_0$ equations~(\ref{Kasner-trans-sys}) can be written in the form:
\eq{t^{2}\sum\limits_{\sss i\ne <n>}\ddot{\varphi}_i=-\psi_{\sss 2n-1,\,2n}^2t^{2-2(p_{2n-1}+p_{2n})}+\sum\limits_{j=\overline{1,\chi} \atop j\ne n}\psi_{\sss 2j-1,\,2j}^2t^{2-2(p_{2j-1}+p_{2j})},\;n=\overline{1,\chi}}
or, in matrix notations,
\eq{A\Phi(t)=R(t),\quad t\in(t_0-\varepsilon,t_0+\varepsilon)}
where
\eq{A_{k}^{k}=0,\quad A_{j}^k=1,\; k\ne j;\quad \Phi_k(t)=\ddot{\varphi}_k(t),\quad j,k=\overline{1,N}}
\eq{R_{2n-1}(t)=R_{2n}(t)=\psi_{\sss 2n-1,\,2n}^2t^{-2(p_{2n-1}+p_{2n})}-\sum\limits_{j=\overline{1,\chi} \atop j\ne n}\psi_{\sss 2j-1,\,2j}^2t^{-2(p_{2j-1}+p_{2j})},\quad n=\overline{1,\chi}}
It is easy to check that matrix $A$ is nonsingular, so we obtain:
\eq{\Phi(t)=A^{-1}R(t),}
or,
\eq{\ddot{\varphi}_k(t)=\sum\limits_{m=\overline{1,\chi}}\eta_{km}t^{\lambda_m},\quad \eta_{km}=\const,\quad \lambda_m=-2(p_{2m-1}+p_{2m}),\quad k=\overline{1,N}}
Hence we deduce that
\eq{\dot{\varphi}_k(t)=\sum\limits_{m=\overline{1,\chi}}\frac{\eta_{km}}{\lambda_m+1}t^{\lambda_m+1},\quad \varphi_k(t)=\sum\limits_{m=\overline{1,\chi}}\frac{\eta_{km}}{\lambda_m+2}t^{\lambda_m+2},\quad k\oneN}
It follows from $p_{2m-1}+p_{2m}<1,\;m=\overline{1,\chi}$ that $\lambda_m+2>0$; therefore
\eq{\lim\limits_{t\rightarrow0}\varphi_k(t)=0,\quad k\oneN}
and solutions~(\ref{Kasner-sol}) with $p_{2m-1}+p_{2m}<1,\;m=\overline{1,\chi}$ are asymptotically stable.\pff\newpage

\noindent \textbf{Examples.}
\begin{itemize}
  \item (3+1)-dimensional space-time; in view of~(\ref{chi-odd}) we have a single non-zero component of the Faraday tensor; according to our convention (see~Sec.\ref{preliminaries}) we denote this component by $\psi_{12}$; under the Proposition 1 it follows that stable Kasner solutions are characterized by the condition $p_1+p_2<1$; it implies that $p_3>0$ --- the space expands along the magnetic field.
  \item (4+1)-dimensional space-time; in that case we have a couple non-zero components of the Faraday tensor: $\psi_{12}$ and $\psi_{34}$; in accordance with the Proposition 1 it follows that stable Kasner solutions are specified by the criteria $p_1+p_2<1$ and $p_3+p_4<1$.
\end{itemize}
\subsection{Numerical calculations.\label{num-Kasner}}
\begin{figure}[t]
\begin{minipage}[h]{.5\linewidth}
\center{\includegraphics[width=.7\linewidth]{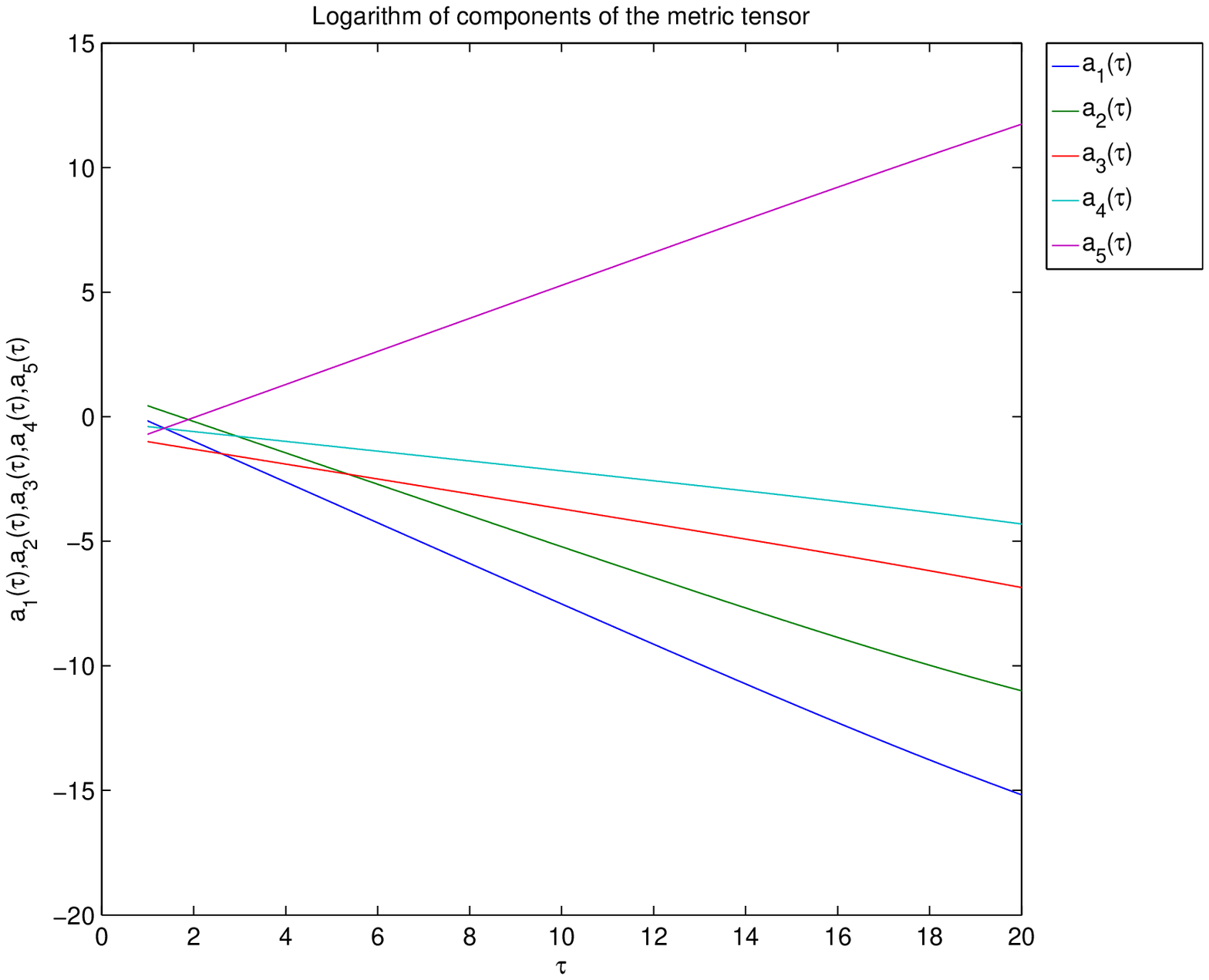} \\ a)}
\end{minipage}
\hfill
\begin{minipage}[h]{.5\linewidth}
\center{\includegraphics[width=.7\linewidth]{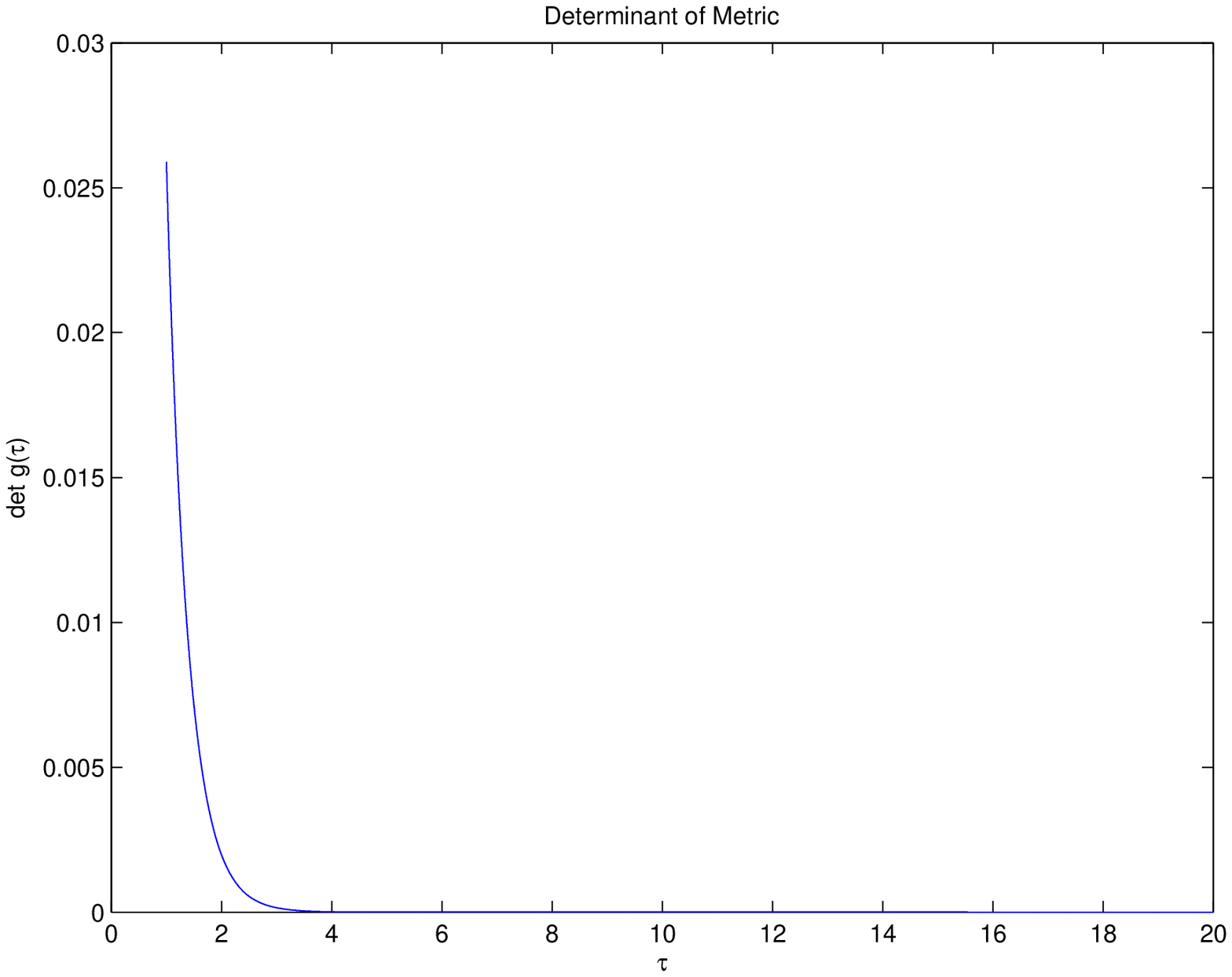} \\ b)}
\end{minipage}
\caption{\footnotesize Numerical solutions for (5+1)-dimension space-time (Einstein gravity). The figure a) illustrates stability of the Kasner solution near the initial singularity. The figure b) shows decreasing of the determinant of the metric; it implies that any given element of volume $dV=\sqrt{|\det(g)|}d^Nx$ tend to zero and therefore we go to the initial singularity.}
\label{review-Kasner}
\end{figure}

The result presented above have been verified numerically. We have got a few hundreds numerical solutions with a random sets of initial conditions for each of the dimensions $N=5,6,7,8$ and found a lot of solutions that converge to the Kasner ones. Example of such solution is presented on the Fig.~\ref{review-Kasner} (a).

It should be note that looking for asymptotically stable solutions numerically using the synchronous time $t$ has a number of drawbacks; particularly, numerical solutions of the equations~(\ref{n-mod5-2})-(\ref{R0-mod}) increase too rapidly impeding computations as $t$ goes to zero. To avoid these problem we have introduced new time coordinate $\tau$ by means of the relation
\eq{d\tau=\e^{-\sum\limits_{j}a_j(t)}dt\label{t1-Kasner}}
The change of variables from $t$ to $\tau$ results in transformation of logarithmic functions~(\ref{Kasner-sol}) into linear ones. Indeed, (\ref{Kasner-sol}) and (\ref{t1-Kasner}) leads to
\eq{d\tau=\frac{C}{t}dt,\quad C=\e^{-\sum\limits_{j}C_j}=\const}
Consequently,
\eq{\tau=C\ln(t)+\widetilde{C},\quad \widetilde{C}=\const}
\eq{a_k(\tau)=\frac{p_k}{C}\tau+C_k-p_k\frac{\widetilde{C}}{C},\quad k\oneN}
Without lack of generality we can set
\eq{C_1=\ldots=C_N=\widetilde{C}=0}
Then, as was announced,
\eq{a_k(\tau)=p_k\tau,\quad k\oneN \label{ak}}
Such a way, looking for asymptotically stable solutions numerically, we should expect to get quasi-linear solutions like ones that shown on the Fig.~\ref{review-Kasner} (a).
\section{Stability of power-law solutions in Gauss-Bonnet models.\label{4}}
\subsection{Field equations.}
Action of the theory under consideration is given by:
\eq{S=\frac{1}{16\pi}\int d^{N+1}x\sqrt{|\det(g)|}\bigl(\mathcal{L}_{GB}+\mathcal{L}_m\bigr)}
\eq{\mathcal{L}_{GB}=R_{\alpha\beta\gamma\delta}R^{\alpha\beta\gamma\delta}-
4R_{\alpha\beta}R^{\alpha\beta}+R^2,\quad\mathcal{L}_m=F_{\alpha\beta}F^{\alpha\beta}}
where $R,R_{\alpha\beta},R_{\alpha\beta\gamma\delta},F_{\alpha\beta}$ are the $(N+1)$-dimensional scalar curvature, Ricci tensor, Riemann tensor and  Faraday tensor respectively. The gravitational equations has the form:
\eq{H^{\mu}_{\nu}=8\pi T^{\mu}_{\nu}\,,\label{eqmot}} where \eq{H^{\mu}_{\nu}=2RR^{\mu}_{\nu}-4R^{\mu}_{\gamma}R^{\gamma}_{\nu}-4R^{\alpha\beta}R^{\mu}_{.\;\alpha\nu\beta}+
2R^{\mu\alpha\beta\gamma}R_{\nu\alpha\beta\gamma}-
\frac{1}{2}\delta^{\mu}_{\nu}\Bigl(R_{\alpha\beta\gamma\delta}R^{\alpha\beta\gamma\delta}-
4R_{\alpha\beta}R^{\alpha\beta}+R^2\Bigr)\label{GB-tensor}}
are the components of the Gauss-Bonnet tensor. Taking into account~(\ref{g}) we get:
\eq{H^{\mu}_{\nu}=0,\;\;\mu\ne\nu\label{Hmn}}
\eq{H^0_0=-12\sum\limits_{i<j<k<l}\dot{a}_i\dot{a}_j\dot{a}_k\dot{a}_l,\quad H^n_n=-4\sum\limits_{i\ne n}\bigl(\ddot{a}_i+\dot{a}_i^2\bigr)\sum\limits_{j<k \atop j,k\ne i,n }\dot{a}_j\dot{a}_k-
12\sum\limits_{i<j<k<l \atop i,j,k,l\ne n}\dot{a}_i\dot{a}_j\dot{a}_k\dot{a}_l,\;\; n\oneN\label{H00-Hnn}}
For the diagonal case  in view of~(\ref{Tmunu}),(\ref{T00-Tnn}),(\ref{eqmot}),(\ref{H00-Hnn}) for an even number of dimensions the gravitational equations can be written as:
\eqs{\sum\limits_{\sss i\ne <n>}\bigl(\ddot{a}_i+\dot{a}_i^2\bigr)\sum\limits_{\sss j<k \atop j,k\ne i,<n>}\dot{a}_j\dot{a}_k+
3\sum\limits_{i<j<k<l \atop i,j,k,l\ne <n>}\dot{a}_i\dot{a}_j\dot{a}_k\dot{a}_l=}
\eq{=\frac{1}{4}\left[\psi_{\sss 2n-1,\,2n}^2e^{-2(a_{2n-1}+a_{2n})}-
\sum\limits_{j=\overline{1,\chi} \atop j\ne n}\psi_{\sss 2j-1,\,2j}^2e^{-2(a_{2j-1}+a_{2j})}\right],\; n=\overline{1,\chi}\label{n-mod5-2}}
\eq{\sum\limits_{i<j<k<l}\dot{a}_i\dot{a}_j\dot{a}_k\dot{a}_l+\frac{1}{12}\sum\limits_{i=\overline{1,\chi}}\psi_{\scriptscriptstyle 2i-1,\,2i}^2e^{-2(a_{2i-1}+a_{2i})}=0,\label{R0-mod}}
There is one more equation in addition to these in the case of an odd number of dimensions:
\eq{\sum\limits_{\sss i\ne N}\bigl(\ddot{a}_i+\dot{a}_i^2\bigr)\sum\limits_{\sss j<k \atop j,k\ne i,N}\dot{a}_j\dot{a}_k+
3\sum\limits_{i<j<k<l \atop i,j,k,l\ne N}\dot{a}_i\dot{a}_j\dot{a}_k\dot{a}_l=-\frac{1}{4}\sum\limits_{i=\overline{1,\chi}}\psi_{\scriptscriptstyle 2i-1,\,2i}^2e^{-2(a_{2i-1}+a_{2i})}\label{n-mod5-1}}
The left side of the expression~(\ref{R0-mod}) is a first integral of~(\ref{n-mod5-2}).
\subsection{Vacuum solution.}
When there is no any matter the equations~(\ref{n-mod5-2})-(\ref{n-mod5-1}) lead to:
\eq{\sum\limits_{\sss i\ne n}\bigl(\ddot{a}_i+\dot{a}_i^2\bigr)\sum\limits_{\sss j<k \atop j,k\ne i,n}\dot{a}_j\dot{a}_k+
3\sum\limits_{i<j<k<l \atop i,j,k,l\ne n}\dot{a}_i\dot{a}_j\dot{a}_k\dot{a}_l=0,\quad n\oneN\label{vac-1}}
\eq{\sum\limits_{i<j<k<l}\dot{a}_i\dot{a}_j\dot{a}_k\dot{a}_l=0\label{vac-2}}
These equations has the following solutions~\cite{Deruelle1,Deruelle2}:
\eq{a_k=p_k\ln(t)+C_k,\quad C_k=\const,\; k\oneN,\qquad \sum\limits_{i<j<l<m}p_i p_j p_l p_m=0,\quad \sum\limits_{n}p_n=3\label{metric-1}}
As a consequence, components of the metric tensor follows a power law:
\eq{g_{kk}(t)=\e^{2a_k(t)}=\overline{C}_kt^{2p_k},\quad \overline{C}_k=\const,\quad k\oneN\label{Kasner-like-metr}}
Thus by analogy with~(\ref{Kasner-sol})-(\ref{Kasner-metr}) we will call solutions~(\ref{metric-1}) and metric~(\ref{Kasner-like-metr}) Kasner-like for brevity.
\subsection{Stability conditions.}
As well as in the case of the Einstein gravity (see~\ref{Einstein-stab.cond}), metric ceases to obey a power law when there is a matter; it appears however that the metric of the space filled with the magnetic field may get close to the Kasner-like metric~(\ref{Kasner-like-metr}) and converge to it when moving towards the initial singularity. Namely, numerical calculations shows that there exists solutions of the equations~(\ref{n-mod5-2}) that get close to the Kasner-like solutions~(\ref{metric-1}) and converge to them as time goes to the point $t=0$. We will looking for solutions of the equations~(\ref{n-mod5-2}) in the form:
\eq{a_k(t)=a_k^0(t)+\varphi_k(t),\quad k\oneN\label{a-a0}}
where $a^0_k$ is the Kasner-like solution~(\ref{metric-1}), $\varphi_k\in C^2(\mathbb{R})$. We will say that solutions $a_k^0$ are asymptotically stable if
\eq{\lim\limits_{t\rightarrow0}a_k(t)=a_k^0(t),\quad k\oneN}
In other words, solutions $a_k^0$ are asymptotically stable if and only if
\eq{\lim\limits_{t\rightarrow0}\varphi_k(t)=0,\quad k\oneN}
Now we will find out conditions which specify asymptotically stable Kasner-like solutions.

\noindent\textbf{Proposition 2}. \emph{Kasner-like solutions are asymptotically stable for $t\rightarrow0$ if and only if
$p_{2n-1}+p_{2n}<\nolinebreak2,\;n=\nolinebreak\overline{1,\chi}$.}\vspace{.2cm}
\pfi We consider the space of an even dimension; that of an odd dimension is treated similar way.\vspace{.1cm}

\noindent 1. Let $\{a_1^0,\ldots,a_N^0\}$ be an asymptotically stable solution of the equations~(\ref{vac-1})-(\ref{vac-2}) given by~(\ref{metric-1}); then there exists solution $\{a_1,\ldots,a_N\}$ of the equations~(\ref{n-mod5-2}),(\ref{R0-mod}) and functions $\varphi_1,\ldots,\varphi_N\in C^2(\mathbb{R})$ such that
\eq{a_k(t)=a_k^0(t)+\varphi_k(t),\quad k\oneN\label{disturbed}}
\eq{\lim\limits_{t\rightarrow0}\varphi_k(t)=0,\quad k\oneN\label{phi}}
Let us introduce the following notation:
\eq{\mathcal{A}_{\textbf{(}i_1}\mathcal{B}_{i_2}\cdot\ldots\cdot\mathcal{B}_{i_N\textbf{)}}=
\mathcal{A}_{i_1}\mathcal{B}_{i_2}\cdot\ldots\cdot\mathcal{B}_{i_N}+\mathcal{A}_{i_2}\mathcal{B}_{i_1}\cdot\ldots\cdot\mathcal{B}_{i_N}+\ldots+
\mathcal{A}_{i_N}\mathcal{B}_{i_2}\cdot\ldots\cdot\mathcal{B}_{i_1}}
where $\mathcal{A}_{i_1},\ldots,\mathcal{A}_{i_N},\mathcal{B}_{i_1},\ldots,\mathcal{B}_{i_N}$ are any indexed mathematical objects.
Substitution~(\ref{metric-1}),(\ref{disturbed}) to the equations~(\ref{n-mod5-2}) and~(\ref{R0-mod}) leads to:
\eqs{\sum\limits_{\sss i\ne <n>}p_i(p_i-1)\sum\limits_{\sss j<k \atop j,k\ne i,<n>}p_j p_k+3\sum\limits_{i<j<k<l \atop i,j,k,l\ne <n>}p_i p_j p_k p_l+}
\eqs{+t\left[\sum\limits_{\sss i\ne <n>}p_i(p_i-1)\sum\limits_{\sss j<k \atop j,k\ne i,<n>}p_{\textbf{(}j}\dot{\varphi}_{k\textbf{)}}+2\sum\limits_{\sss i\ne <n>}p_i\dot{\varphi}_i\sum\limits_{\sss j<k \atop j,k\ne i,<n>}p_j p_k+3\sum\limits_{i<j<k<l \atop i,j,k,l\ne <n>}\dot{\varphi}_{\textbf{(}i} p_j p_k p_{l\textbf{)}}\right]+}
\eqs{+t^{2}\left[\sum\limits_{\sss i\ne <n>}p_i(p_i-1)\sum\limits_{\sss j<k \atop j,k\ne i,<n>}\dot{\varphi}_j \dot{\varphi}_k+2\sum\limits_{\sss i\ne <n>}p_i\dot{\varphi}_i\sum\limits_{\sss j<k \atop j,k\ne i,<n>}p_{\textbf{(}j}\dot{\varphi}_{k\textbf{)}}\right]+}
\eqs{t^{2}\left[\sum\limits_{\sss i\ne <n>}\Bigl(\ddot{\varphi}_i+\dot{\varphi}_i^2\Bigr)\sum\limits_{\sss j<k \atop j,k\ne i,<n>}p_j p_k+3\sum\limits_{i<j<k<l \atop i,j,k,l\ne <n>}\Bigl(\dot{\varphi}_{\textbf{(}i} p_k p_{l\textbf{)}} \dot{\varphi}_j+\dot{\varphi}_{i}\dot{\varphi}_{\textbf{(}j} p_k p_{l\textbf{)}}\Bigr)\right]+}
\eqs{+t^{3}\left[2\sum\limits_{\sss i\ne <n>}p_i\dot{\varphi}_i\sum\limits_{\sss j<k \atop j,k\ne i,<n>}\dot{\varphi}_j \dot{\varphi}_k+\sum\limits_{\sss i\ne <n>}\Bigl(\ddot{\varphi}_i+\dot{\varphi}_i^2\Bigr)\sum\limits_{\sss j<k \atop j,k\ne i,<n>}p_{\textbf{(}j}\dot{\varphi}_{k\textbf{)}}+3\sum\limits_{i<j<k<l \atop i,j,k,l\ne <n>}p_{\textbf{(}i} \dot{\varphi}_j \dot{\varphi}_k \dot{\varphi}_{l\textbf{)}}\right]+}
\eqs{+t^4\left[\sum\limits_{\sss i\ne <n>}\Bigl(\ddot{\varphi}_i+\dot{\varphi}_i^2\Bigr)\sum\limits_{\sss j<k \atop j,k\ne i,<n>}\dot{\varphi}_j \dot{\varphi}_k+3\sum\limits_{i<j<k<l \atop i,j,k,l\ne <n>}\dot{\varphi}_{i} \dot{\varphi}_j \dot{\varphi}_k \dot{\varphi}_{l}\right]=}
\eq{=\frac{1}{4}\left[\psi_{\sss 2n-1,\,2n}^2e^{-2(\varphi_{2n-1}+\varphi_{2n})}t^{4-2(p_{2n-1}+p_{2n})}-
\sum\limits_{j=\overline{1,\chi} \atop j\ne n}\psi_{\sss 2j-1,\,2j}^2e^{-2(\varphi_{2j-1}+\varphi_{2j})}t^{4-2(p_{2j-1}+p_{2j})}\right]\label{trans-sys}}
\eqs{\sum\limits_{i<j<k<l \atop i,j,k,l\ne <n>}p_i p_j p_k p_l+t\sum\limits_{i<j<k<l \atop i,j,k,l\ne <n>}\dot{\varphi}_{\textbf{(}i} p_j p_k p_{l\textbf{)}}+t^2\sum\limits_{i<j<k<l \atop i,j,k,l\ne <n>}\Bigl(\dot{\varphi}_{\textbf{(}i} p_k p_{l\textbf{)}} \dot{\varphi}_j+\dot{\varphi}_{i}\dot{\varphi}_{\textbf{(}j} p_k p_{l\textbf{)}}\Bigr)+}
\eq{+t^3\sum\limits_{i<j<k<l \atop i,j,k,l\ne <n>}p_{\textbf{(}i} \dot{\varphi}_j \dot{\varphi}_k \dot{\varphi}_{l\textbf{)}}+t^4\sum\limits_{i<j<k<l \atop i,j,k,l\ne <n>}\dot{\varphi}_{i} \dot{\varphi}_j \dot{\varphi}_k \dot{\varphi}_{l}=-\frac{1}{12}\sum\limits_{i=\overline{1,\chi}}\psi_{\scriptscriptstyle 2i-1,\,2i}^2e^{-2(\varphi_{2j-1}+\varphi_{2j})}t^{4-2(p_{2i-1}+p_{2i})}\label{trans-constr}}
It follows from~(\ref{phi}) that
\eq{\lim\limits_{t\rightarrow0}\bigl(t\dot{\varphi}_k(t)\bigr)=0,\quad\lim\limits_{t\rightarrow0}\bigl(t^2\ddot{\varphi}_k(t)\bigr)=0,\quad k\oneN\label{phi-dot}}
In the limit $t\rightarrow0$ in view of~(\ref{metric-1}),(\ref{phi}),(\ref{phi-dot}) equations~(\ref{trans-sys})-(\ref{trans-constr}) take the form:
\eq{0=\frac{1}{4}\lim\limits_{t\rightarrow0}\left[\psi_{\sss 2n-1,\,2n}^2t^{4-2(p_{2n-1}+p_{2n})}-
\sum\limits_{j=\overline{1,\chi} \atop j\ne n}\psi_{\sss 2j-1,\,2j}^2t^{4-2(p_{2j-1}+p_{2j})}\right]\label{lim-sys}}
\eq{0=-\frac{1}{12}\lim\limits_{t\rightarrow0}\sum\limits_{i=\overline{1,\chi}}\psi_{\scriptscriptstyle 2i-1,\,2i}^2t^{4-2(p_{2i-1}+p_{2i})}\label{lim-constr}}
Equalities~(\ref{lim-sys}) and~(\ref{lim-constr}) are satisfied if and only if
\eq{p_{2n-1}+p_{2n}<2,\quad n=\overline{1,\chi}\label{ineq}}
q.e.d.

2. Let $\{a_1^0,\ldots,a_N^0\}$ be the Kasner-like solution given by~(\ref{metric-1}) such that $p_{2n-1}+p_{2n}<2,\;n=\overline{1,\chi}$, and $\varphi_1(t),\ldots ,\varphi_N(t)\in C^2(\mathbb{R})$ be a small deviations from that solution. We have:
\eq{a_k(t)=a_k^0(t)+\varphi_k(t),\quad k\oneN\label{ak-mod-1}}
Let $t=t_0>0$ be the initial moment; we assume that
\eq{|\varphi_i(t_0)|\ll |a_k^0(t_0)|,\quad |\dot{\varphi}_i(t_0)|\ll \left|\dot{a}_k^0(t_0)\right|,\quad i\oneN\label{dev}}
Substitution~(\ref{metric-1}),(\ref{ak-mod-1}) to the equations~(\ref{n-mod5-2}) and~(\ref{R0-mod}) give us equations~(\ref{trans-sys})-(\ref{trans-constr}). In view of~(\ref{dev}) we can neglect terms contained factors like $t\dot{\varphi}_i,\;t^2\dot{\varphi}_i\dot{\varphi}_j,\;t^3\dot{\varphi}_{i}\dot{\varphi}_j \dot{\varphi}_k,\;t^4\dot{\varphi}_{i}\dot{\varphi}_j\dot{\varphi}_k\dot{\varphi}_{l}$ in the lhs of the equations~(\ref{trans-sys})-(\ref{trans-constr}) and terms contained $\varphi_i$ in the rhs of these equations. Namely, let $\varepsilon$ be a small positive real number; then in the $\varepsilon$-vicinity of the point $t_0$ equations~(\ref{trans-sys}) can be written in the form:
\eq{t^{2}\sum\limits_{\sss i\ne <n>}\ddot{\varphi}_i\sum\limits_{\sss j<k \atop j,k\ne i,<n>}p_j p_k=\frac{1}{4}\left[\psi_{\sss 2n-1,\,2n}^2t^{4-2(p_{2n-1}+p_{2n})}-\sum\limits_{j=\overline{1,\chi} \atop j\ne n}\psi_{\sss 2j-1,\,2j}^2t^{4-2(p_{2j-1}+p_{2j})}\right],\;n=\overline{1,\chi}}
or, in matrix notations,
\eq{A\Phi(t)=R(t),\quad t\in(t_0-\varepsilon,t_0+\varepsilon)}
where
\eq{A_{2n-1}^k=\sum\limits_{i<j \atop i,j\ne k\ne 2n-1}p_i p_j,\quad A_{2n}^k=\sum\limits_{i<j \atop i,j\ne k\ne 2n}p_i p_j,\quad \Phi_k(t)=\ddot{\varphi}_k(t),\quad n=\overline{1,\chi},\;k=\overline{1,N}} \eq{R_{2n-1}(t)=R_{2n}(t)=\frac{1}{4}\left[\psi_{\sss 2n-1,\,2n}^2t^{2-2(p_{2n-1}+p_{2n})}-\sum\limits_{j=\overline{1,\chi} \atop j\ne n}\psi_{\sss 2j-1,\,2j}^2t^{2-2(p_{2j-1}+p_{2j})}\right],\quad n=\overline{1,\chi}}
Assuming that matrix $A$ is nonsingular\footnote{Matrix is singular iff its determinant is 0. Equation $\det\bigl(A(p_1,\ldots,p_N)\bigr)=0$ specifies a surface in $(N-2)$-dimensional space of exponents $p_k$ that determine Kasner-like solutions (we have $N$ parameters $p_1,\ldots,p_N$ connected by two relations: see~(\ref{metric-1})). Thus, excluding that surface from our consideration we lose no more than set of measure zero; as a consequence, our assumption has no significant effect on interesting result.}, we obtain:
\eq{\Phi(t)=A^{-1}R(t),}
or,
\eq{\ddot{\varphi}_k(t)=\sum\limits_{m=\overline{1,\chi}}\delta_{km}t^{\gamma_m},\quad \delta_{km}=\const,\quad \gamma_m=2-2(p_{2m-1}+p_{2m}),\quad k=\overline{1,N}}
Hence we deduce that
\eq{\dot{\varphi}_k(t)=\sum\limits_{m=\overline{1,\chi}}\frac{\delta_{km}}{\gamma_m+1}t^{\gamma_m+1},\quad \varphi_k(t)=\sum\limits_{m=\overline{1,\chi}}\frac{\delta_{km}}{\gamma_m+2}t^{\gamma_m+2},\quad k\oneN}
It follows from $p_{2m-1}+p_{2m}<2,\;m=\overline{1,\chi}$ that $\gamma_m+2>0$; therefore
\eq{\lim\limits_{t\rightarrow0}\varphi_k(t)=0,\quad k\oneN}
and solutions~(\ref{metric-1}) with $p_{2m-1}+p_{2m}<2,\;m=\overline{1,\chi}$ are asymptotically stable.\pff

\noindent \textbf{Example.} Let us consider (5+1)-dimensional space-time. According to~(\ref{chi-odd}) there are pair of non-zero components of the Faraday tensor; taking into account our convention (see~\ref{preliminaries} for details) we denote them $\psi_{12}$ and $\psi_{34}$. Then under the Proposition 2 it follows that stable Kasner-like solutions are described by the conditions $p_1+p_2<2$ and $p_3+p_4<2$.

\subsection{Numerical calculations.\label{Kasner-like.num}}
\begin{figure}[t]
\begin{minipage}[h]{.5\linewidth}
\center{\includegraphics[width=.7\linewidth]{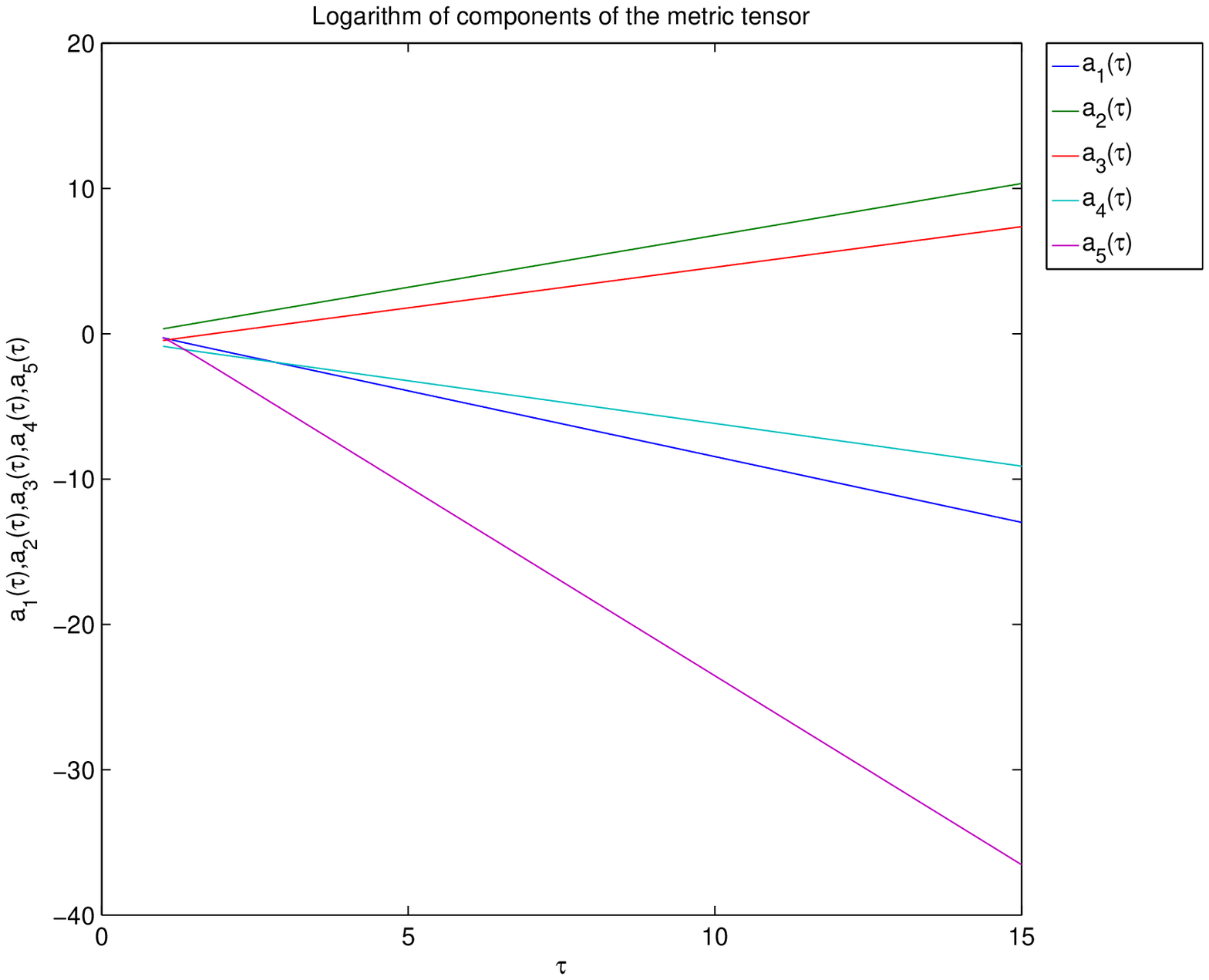} \\ a)}
\end{minipage}
\hfill
\begin{minipage}[h]{.5\linewidth}
\center{\includegraphics[width=.7\linewidth]{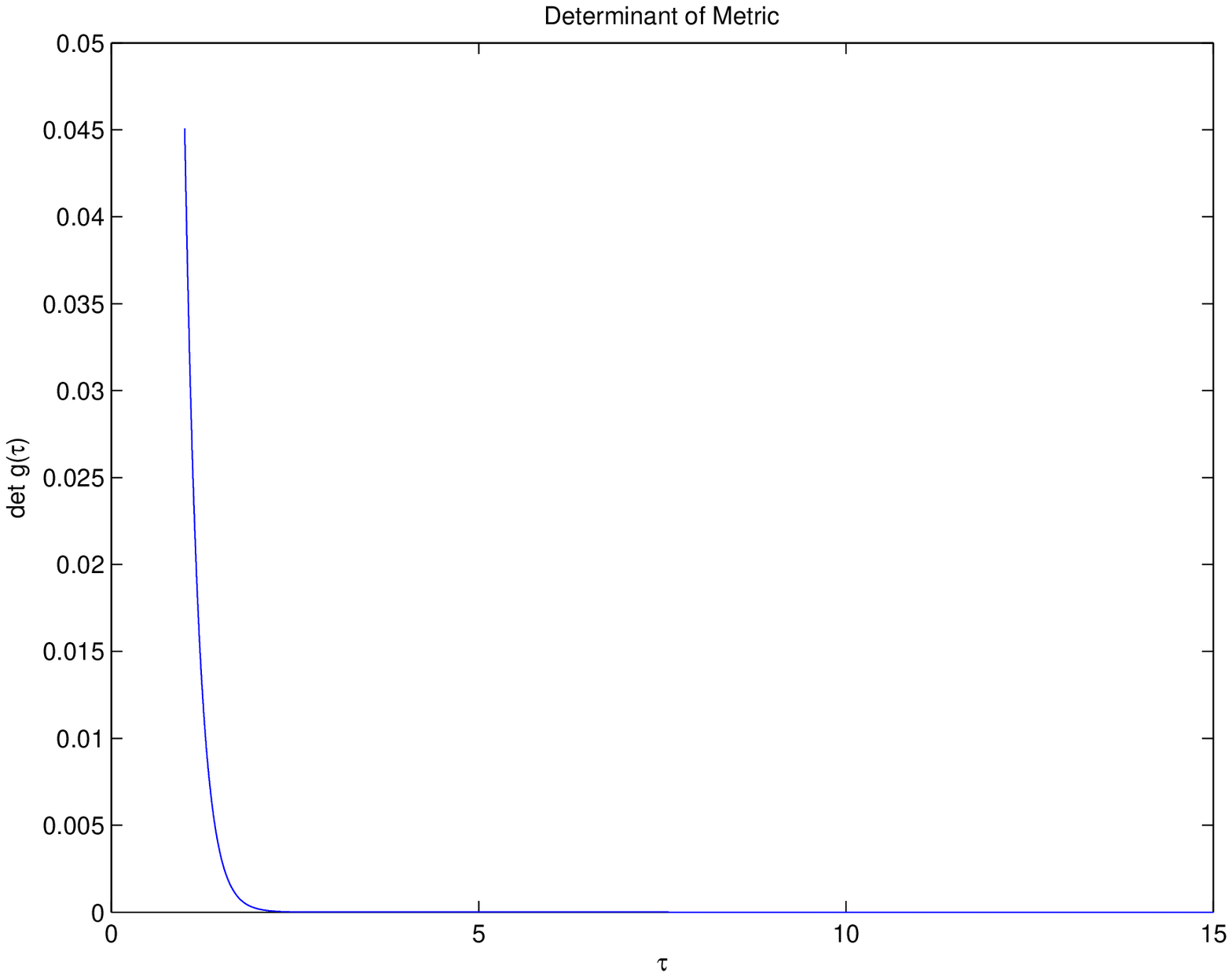} \\ b)}
\end{minipage}
  \caption{\footnotesize Numerical solutions for (5+1)-dimension space-time (Gauss-Bonnet gravity). The figure a) illustrates stability of the Kasner-like regime near the initial singularity. The figure b) shows decreasing of the determinant of the metric; it implies that any given element of volume $dV=\sqrt{|\det(g)|}d^Nx$ tend to zero and, consequently, we go to the initial singularity.}\label{review-1}
\end{figure}
The  result obtained have been  confirmed by numerical calculations. As well as in the study of the stability of the Kasner solutions (see~\ref{Einstein-stab.cond}) we have received a few hundreds solutions numerically with a random sets of initial conditions for each of the dimensions $N=5,6,7,8$ and found numerous solutions that converge to the Kasner-like ones. Example of such solution is presented on the Fig.~\ref{review-1} (a).

For the reasons mentioned above (see~\ref{num-Kasner}) one should use special time coordinate for numerical calculations. Let us introduce new time coordinate $\tau$ by the following way:
\eq{d\tau=\e^{-\frac{1}{3}\sum\limits_{j}a_j(t)}dt\label{t1}}
The change of variables from $t$ to $\tau$ results in transformation of logarithmic functions~(\ref{metric-1}) into linear ones. Indeed, (\ref{metric-1}) and (\ref{t1}) give us:
\eq{d\tau=\frac{C}{t}dt,\quad C=\e^{-\frac{1}{3}\sum\limits_{j}C_j}=\const}
Consequently,
\eq{\tau=C\ln(t)+\widetilde{C},\quad \widetilde{C}=\const}
\eq{a_k(\tau)=\frac{p_k}{C}\tau+C_k-p_k\frac{\widetilde{C}}{C},\quad k\oneN}
Without loss of generality we can set
\eq{C_1=\ldots=C_N=\widetilde{C}=0}
Then, as was announced,
\eq{a_k(\tau)=p_k\tau,\quad k\oneN \label{ak}}
Thus, searching for asymptotically stable solutions numerically we should expect to get quasi-linear solutions like ones that shown on the Fig.~\ref{review-1} (a).

\section{Oscillatory regime.\label{5}}
\begin{figure}[!h]
\begin{minipage}[h]{\linewidth}
\center{\includegraphics[width=.85\linewidth]{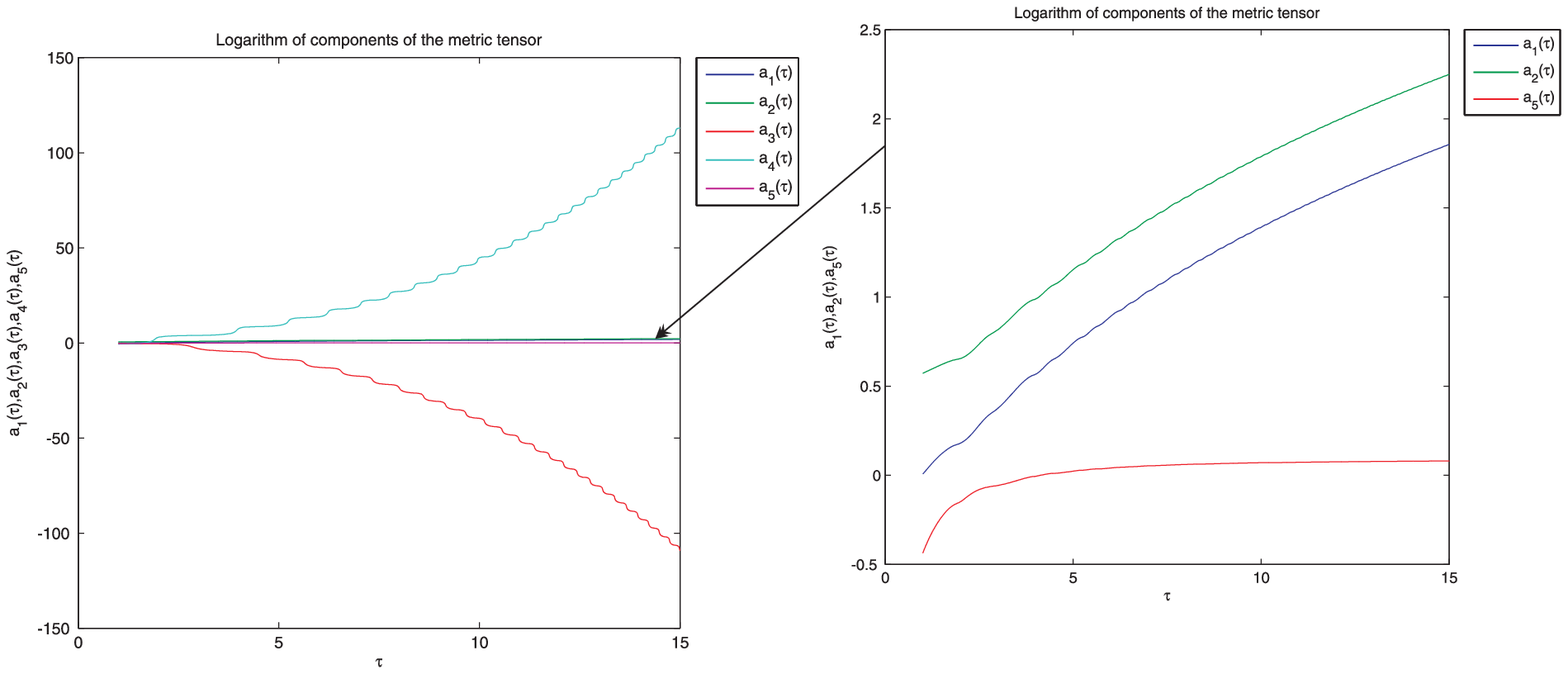} \\ a)}
\end{minipage}
\vfill
\begin{minipage}[h]{\linewidth}
\center{\includegraphics[width=.45\linewidth]{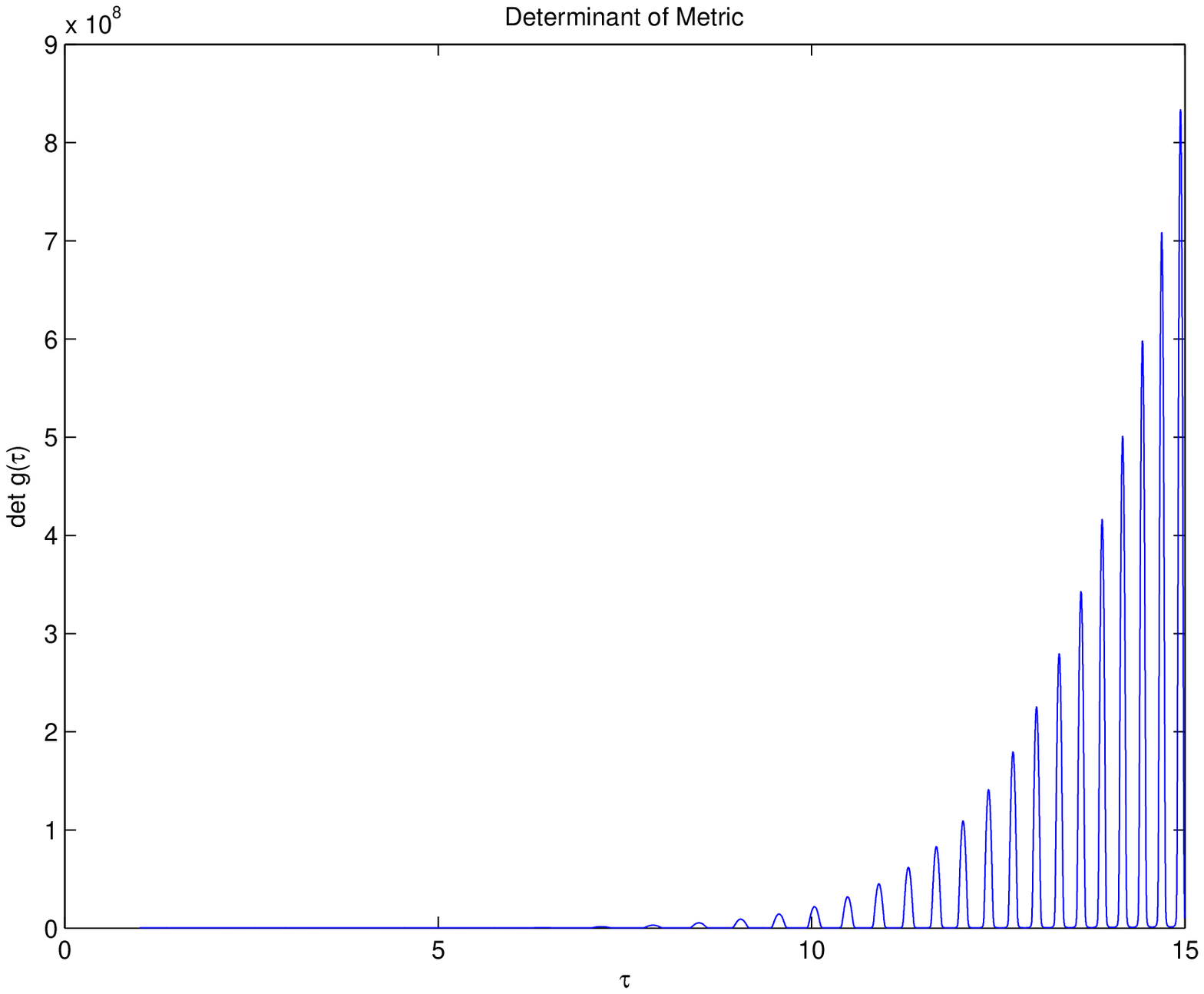} \\ b)}
\end{minipage}
\caption{\footnotesize Numerical solutions for (5+1)-dimension space-time (Gauss-Bonnet gravity). The figure a) depicts oscillatory solution which does not tend to Kasner-like solutions both to the past and to the future. The figure b) shows the oscillatory behavior of the determinant of the metric; it means that any given volume element $dV=\sqrt{|\det(g)|}d^Nx$ oscillates in this regime.}
\label{review-2}
\end{figure}

Results of the preceding section pose an interesting question about the fate of trajectories in the instability zone.
They can either reach stable Kasner-like attractor or represent some other type of dynamics different from Kasner-like
behavior. In this section we show that the latter alternative can be realized.
In our numerical calculations we have revealed a specific class of solutions of the equations~(\ref{n-mod5-2}). The discovered solutions possess the following properties: two of $N$ functions $a_1,\ldots,a_N$ oscillate and has opposite signs, the other ones vary slowly; owing to phase difference between the oscillating functions volume element $\sqrt{|\det(g)|}d^Nx$ oscillates too. In view of limited computational capabilities we have obtained numerical solutions of the equations~(\ref{n-mod5-2}) only for the cases $N=5,6,7,8$; for each of that dimensions we have found out sufficiently large number of oscillatory solutions. Example of such solution is presented on the Fig.~\ref{review-2} (a); the corresponding oscillations of the volume element is shown on the Fig.~\ref{review-2} (b).

\begin{figure}[h]
\begin{minipage}[h]{1.0\linewidth}
\center{\includegraphics[width=1.0\linewidth]{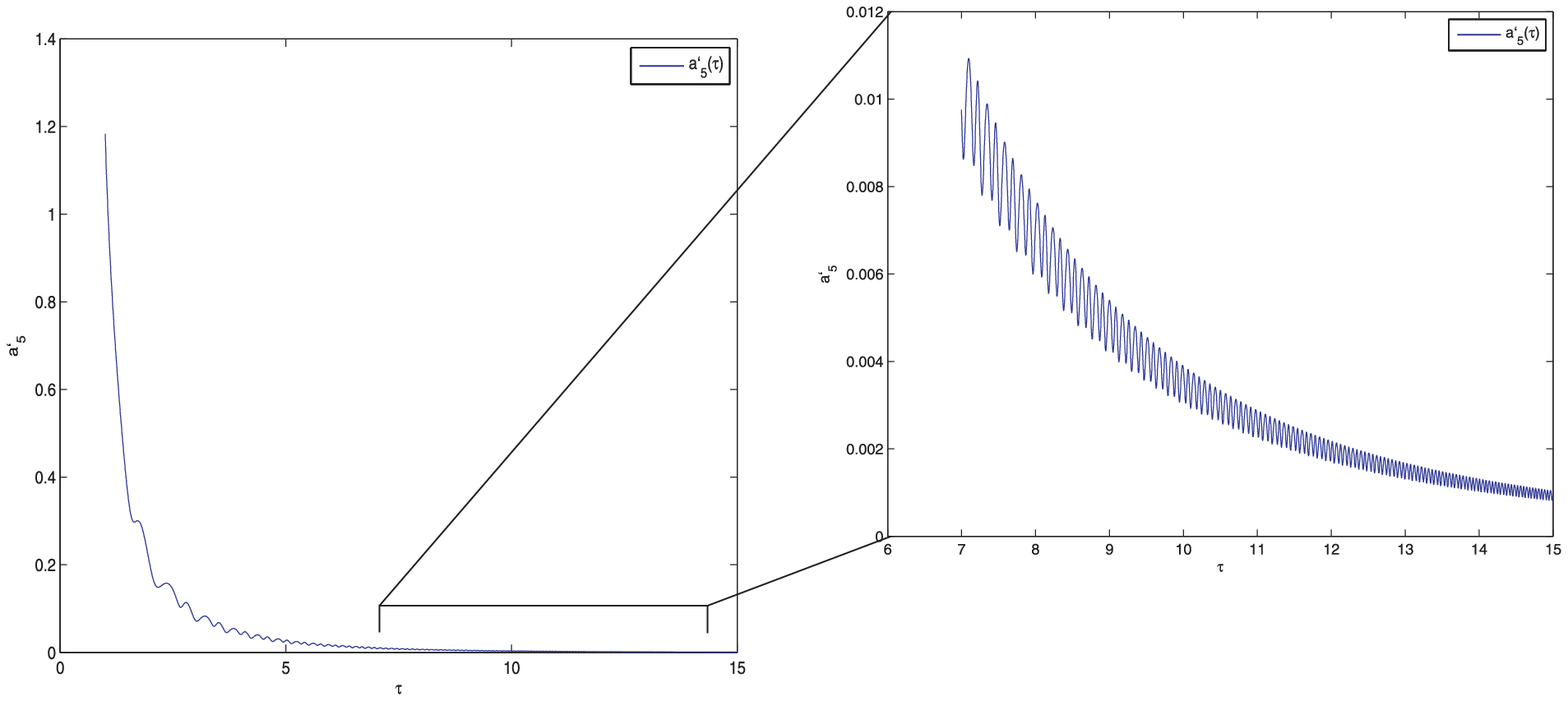} \\ a)}
\end{minipage}
\vfill
\begin{minipage}[h]{1.0\linewidth}
\center{\includegraphics[width=1.0\linewidth]{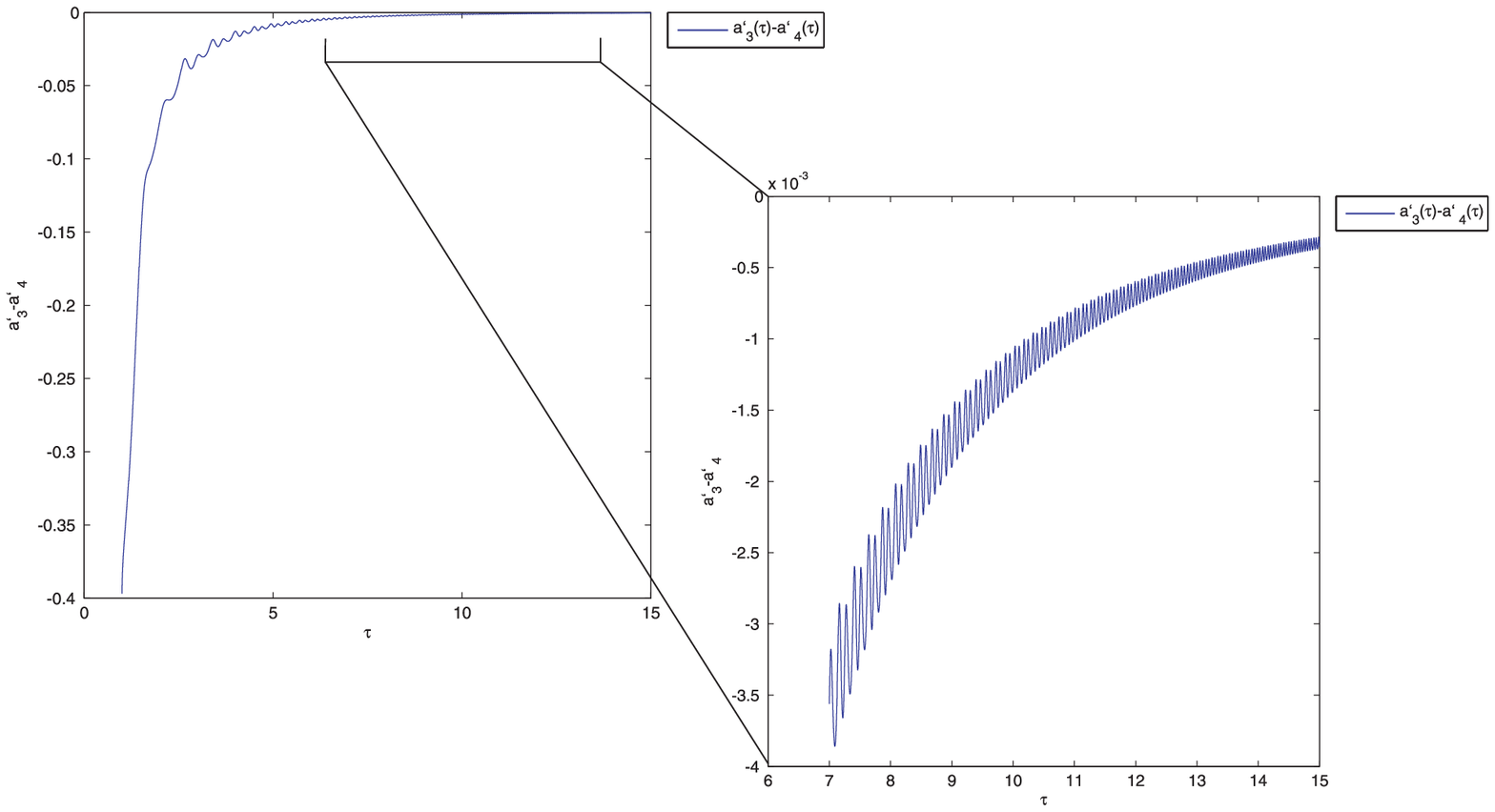} \\ b)}
\end{minipage}
\caption{\footnotesize a) The figure shows that the derivative $a'_5(\tau)$ asymptotically tends to zero. b) The figure indicates that the difference $a'_3(\tau)-a'_4(\tau)$ asymptotically tends to zero.}
\label{num-anlt-5}
\end{figure}

For $N=5$ we have been able to derive an analytical approximation to such solutions on a certain interval of time. Below we describe the procedure of derivation of this approximation. Henceforth we use time coordinate $\tau$ introduced by~(\ref{t1}). In the (5+1)-dimensional case equations~(\ref{n-mod5-2}) become:
\eq{\Bigl(a'_2\bigl[a'_3a'_4+a'_3a'_5+a'_4a'_5\bigr]+a'_3a'_4a'_5\Bigr)'=\frac{1}{6}e^{\frac{4}{3}\sum\limits_{j=1}^5 a_j}\left(\psi_{12}^2e^{-2a_1-2a_2}-2\psi_{34}^2e^{-2a_3-2a_4}\right)\label{d5-1}}
\eq{\Bigl(a'_1\bigl[a'_3a'_4+a'_3a'_5+a'_4a'_5\bigr]+a'_3a'_4a'_5\Bigr)'=\frac{1}{6}e^{\frac{4}{3}\sum\limits_{j=1}^5 a_j}\left(\psi_{12}^2e^{-2a_1-2a_2}-2\psi_{34}^2e^{-2a_3-2a_4}\right)}
\eq{\Bigl(a'_1a'_2\bigl[a'_4+a'_5\bigr]+a'_4a'_5\bigl[a'_1+a'_2\bigr]\Bigr)'=\frac{1}{6}e^{\frac{4}{3}\sum\limits_{j=1}^5 a_j}\left(\psi_{34}^2e^{-2a_3-2a_4}-2\psi_{12}^2e^{-2a_1-2a_2}\right)}
\eq{\Bigl(a'_1a'_2\bigl[a'_3+a'_5\bigr]+a'_3a'_5\bigl[a'_1+a'_2\bigr]\Bigr)'=\frac{1}{6}e^{\frac{4}{3}\sum\limits_{j=1}^5 a_j}\left(\psi_{34}^2e^{-2a_3-2a_4}-2\psi_{12}^2e^{-2a_1-2a_2}\right)}
\eq{\Bigl(a'_1a'_2\bigl[a'_3+a'_4\bigr]+a'_3a'_4\bigl[a'_1+a'_2\bigr]\Bigr)'=-\frac{1}{3}e^{\frac{4}{3}\sum\limits_{j}a_j}
\left(\psi_{12}^2e^{-2a_1-2a_2}+\psi_{34}^2e^{-2a_3-2a_4}\right)\label{d5-5}}
with the first integral
\eq{a'_1 a'_2 a'_3 a'_4+a'_1 a'_2 a'_3 a'_5+a'_1 a'_2 a'_4 a'_5+a'_1 a'_3 a'_4 a'_5+a'_2 a'_3 a'_4 a'_5=-\frac{1}{12}e^{\frac{4}{3}\sum\limits_{j}a_j}\left(\psi_{12}^2e^{-2a_1-2a_2}+\psi_{34}^2e^{-2a_3-2a_4}\right)\label{d5-6}}
The dash denotes derivative with respect to $\tau$. Analysis of numerical solutions shows that there exists $\tau_0>0$ ($\tau=0$ is accepted as the initial moment) such that
\eq{a'_5(\tau)\approx 0,\quad a'_3(\tau)\approx a'_4(\tau),\quad \tau>\tau_0\label{a3a4a5}}
(see Fig.~\ref{num-anlt-5}). In view of~(\ref{a3a4a5}) equations~(\ref{d5-1})-(\ref{d5-6}) take the form:
\eq{\Bigl(a'_1a'_3a'_4\Bigr)'=\frac{f}{2},\quad \Bigl(a'_2a'_3a'_4\Bigr)'=\frac{f}{2},\quad \Bigl(a'_1a'_2a'_3\Bigr)'=\frac{g}{2},\quad \Bigl(a'_1a'_2a'_4\Bigr)'=\frac{g}{2}\label{fg}} \eq{\Bigl(a'_1a'_2\bigl[a'_3+a'_4\bigr]+a'_3a'_4\bigl[a'_1+a'_2\bigr]\Bigr)'=h\label{h}} \eq{a'_1 a'_2 a'_3 a'_4=\frac{h}{4}}
where
\eq{f=\left(\kappa_1 e^{\frac{4}{3}v-\frac{2}{3}u}-2\kappa_2 e^{\frac{4}{3}u-\frac{2}{3}v}\right),\quad g=\left(-2\kappa_1 e^{\frac{4}{3}v-\frac{2}{3}u}+\kappa_2 e^{\frac{4}{3}u-\frac{2}{3}v}\right),\quad h=-\left(\kappa_1 e^{\frac{4}{3}v-\frac{2}{3}u}+\kappa_2 e^{\frac{4}{3}u-\frac{2}{3}v}\right)}
\eq{v=a_3+a_4,\quad u=a_1+a_2} \eq{\kappa_1=\frac{\psi_{12}^2}{3},\quad \kappa_2=\frac{\psi_{34}^2}{3}}
Equation~(\ref{h}) is the sum of the equations~(\ref{fg}); equations~(\ref{fg}) can be replaced by their equivalent:
\eq{\Bigl(\bigl[a'_1+a'_2\bigr]a'_3a'_4\Bigr)'=f,\quad \Bigl(\bigl[a'_1-a'_2\bigr]a'_3a'_4\Bigr)'=0,\quad \Bigl(a'_1a'_2\bigl[a'_3+a'_4\bigr]\Bigr)'=g,\quad \Bigl(a'_1a'_2\bigl[a'_3-a'_4\bigr]\Bigr)'=0\label{+-}}
In view of~(\ref{a3a4a5}) the last equation from~(\ref{+-}) is satisfied automatically, and the other equations become
\eq{\Bigl(u'v'^2\Bigr)'=4f,\quad \Bigl(a'_1a'_2v'\Bigr)'=g,\quad \Bigl(\bigl[a'_1-a'_2\bigr]v'^2\Bigr)'=0\label{fg-mod}}
The last of equations~(\ref{fg-mod}) gives:
\eq{a'_1-a'_2=\frac{C}{v'^2},\quad C=\const} so that
\eq{a'_1=\frac{1}{2}\left(u'+\frac{C}{v'^2}\right),\quad a'_2=\frac{1}{2}\left(u'-\frac{C}{v'^2}\right)\label{a1a2}}
Let us decompose functions $f,g,h$ with respect to $\left(\frac{4}{3}v-\frac{2}{3}u\right)$ and $\left(\frac{4}{3}u-\frac{2}{3}v\right)$ up to the first order:
\eq{f\approx \bigl[\kappa_1-2\kappa_2\bigr]-\frac{2}{3}\bigl[\kappa_1+4\kappa_2\bigr]u+\frac{4}{3}\bigl[\kappa_1+\kappa_2\bigr]v}
\eq{g\approx \bigl[\kappa_2-2\kappa_1\bigr]+\frac{4}{3}\bigl[\kappa_1+\kappa_2\bigr]u-\frac{2}{3}\bigl[\kappa_2+4\kappa_1\bigr]v}
\eq{h\approx -\bigl[\kappa_1+\kappa_2\bigr]+\frac{2}{3}\bigl[\kappa_1-2\kappa_2\bigr]u+\frac{2}{3}\bigl[\kappa_2-2\kappa_1\bigr]u\label{fgh}}
Numerical results show that such assumption is valid in a certain interval of time. Taking into account~(\ref{a1a2})-(\ref{fgh}) we get:
\eq{\Bigl(u'v'^2\Bigr)'=4\bigl[\kappa_1-2\kappa_2\bigr]-\frac{8}{3}\bigl[\kappa_1+4\kappa_2\bigr]u+\frac{16}{3}\bigl[\kappa_1+\kappa_2\bigr]v\label{u}} \eq{\left(\left[(u')^2-\frac{C^2}{v'^4}\right]v'\right)'=
4\bigl[\kappa_2-2\kappa_1\bigr]+\frac{16}{3}\bigl[\kappa_1+\kappa_2\bigr]u-\frac{8}{3}\bigl[\kappa_2+4\kappa_1\bigr]v\label{v-eq}}
Let us consider the equation~(\ref{v-eq}). Numerical calculations show that in the regime under consideration $v''/(v')^4$ is the leading term in the lhs of the equation~(\ref{v-eq}); then:
\eq{\frac{C^2v''}{v'^4}=4\bigl[\kappa_2-2\kappa_1\bigr]+\frac{16}{3}\bigl[\kappa_1+\kappa_2\bigr]u-
\frac{8}{3}\bigl[\kappa_2+4\kappa_1\bigr]v\label{v-eq-mod-2}}
We can obtain analytical solution of the equation~(\ref{v-eq-mod-2}) by omitting the term containing $u$ in the rhs of~(\ref{v-eq-mod-2}):
\eq{\frac{v''}{v'^4}=4\bigl[\kappa_2-2\kappa_1\bigr]-\frac{8}{3}\bigl[\kappa_2+4\kappa_1\bigr]v\label{v-eq-mod}}
Solution of the equation~(\ref{v-eq-mod}) has the form
\eq{v(\tau)=\eta\sqrt{\tau}+\lambda,\quad \eta,\lambda=\const\label{v-sol}}
Let us consider the equation~(\ref{u}). We omit the term, containing $v$ in the rhs of~(\ref{u}):
\eq{\Bigl(u'v'^2\Bigr)'=4\bigl[\kappa_1-2\kappa_2\bigr]-\frac{8}{3}\bigl[\kappa_1+4\kappa_2\bigr]u\label{u1}}
Under substitution of~(\ref{v-sol}) into (\ref{u1}) we obtain:
\eq{\tau u''(\tau)-u'(\tau)+\xi^2\tau^2u(\tau)=\zeta\tau^2\label{u2}}
We have used the following notations:
\eq{\xi^2=\frac{32\bigl[\kappa_1+4\kappa_2\bigr]}{3\eta^2},\quad \zeta=\frac{16\bigl[\kappa_1-2\kappa_2\bigr]}{\eta^2}}
Equation~(\ref{u2}) has the following solution:
\eq{u(\tau)=\tau C_1J_{\frac{2}{3}}\left(\frac{2\xi \tau^{3/2}}{3}\right)+\tau C_2Y_{\frac{2}{3}}\left(\frac{2\xi \tau^{3/2}}{3}\right)+{\frac {\zeta}{{\xi}^{2}}},\quad C_1,C_2=\const}
Here $J_{\frac{2}{3}},Y_{\frac{2}{3}}$ are Bessel functions of the first and the second kind respectively. Going back to the initial variables we get:
\eq{a'_1-a'_2=\frac{C}{v'^2},\quad a_3+a_4=v,\;\;a'_3=a'_4,\quad a'_5=0}
Consequently,
\eq{a_1(\tau)-a_2(\tau)=\frac{2C}{\eta^2}\tau^2+\overline{C},\quad a_3(\tau)-\widetilde{C}=a_4(\tau)=\eta\sqrt{\tau}+\lambda,\quad \overline{C},\widetilde{C}=\const}
and, finally:
\eq{a_1(\tau)\approx-C_0\tau^2+\tau\left[C_1J_{\frac{2}{3}}\left(\frac{2\xi \tau^{3/2}}{3}\right)+C_2Y_{\frac{2}{3}}\left(\frac{2\xi \tau^{3/2}}{3}\right)\right]+D_1\label{f1}}
\eq{a_2(\tau)\approx C_0\tau^2+\tau\left[C_1J_{\frac{2}{3}}\left(\frac{2\xi \tau^{3/2}}{3}\right)+C_2Y_{\frac{2}{3}}\left(\frac{2\xi \tau^{3/2}}{3}\right)\right]+D_2}
\eq{a_3(\tau)\approx\eta\sqrt{\tau}+D_3,\quad a_4(\tau)\approx\eta\sqrt{\tau}+\lambda,\quad a_5(\tau)\approx D \label{f5},}
where
\eq{C_0,D_1,D_2,D_3,D=\const}
To demonstrate oscillations of volume element $\sqrt{|\det(g)|}d^Nx$ one should write down expression for the determinant of the metric:
\eq{|\det(g)|=\e^{2(a_1+\ldots+a_5)}=\exp\left\{4\left(\eta\sqrt{\tau}+\tau\left[C_1J_{\frac{2}{3}}\left(\frac{2\xi \tau^{3/2}}{3}\right)+C_2Y_{\frac{2}{3}}\left(\frac{2\xi \tau^{3/2}}{3}\right)\right]+\overline{D}\right)\right\}}
\eq{\overline{D}=D+D_1+D_2+D_3+\lambda=\const}
Despite the fact that functions~(\ref{f1})-(\ref{f5}) are the solution of the simplified system~(\ref{v-eq-mod}),(\ref{u1}) they catch correctly qualitative features of solution of the full system~(\ref{d5-1})-(\ref{d5-5}) which can be shown by comparing functions~(\ref{f1})-(\ref{f5}) with numerical solution of the system~(\ref{d5-1})-(\ref{d5-5}). Figures~\ref{num-anlt-1},\ref{num-anlt-3} below depicts pairs $(a_2,A_2);(a_4,A_4)$ of numerical and analytical solutions; pairs $(a_1,A_1);(a_3,A_3)$ behave themselves analogously.

This regime is not similar to Kasner-like one (Kasner-like solution has linear behavior on chosen time coordinate $\tau$ (see~\ref{Kasner-like.num} for details), the leading asymptotic of the oscillatory solution has quadratic behaviour w.r.t. $\tau$).

\begin{figure}[h]
\begin{minipage}[h]{1.0\linewidth}
\center{\includegraphics[width=1.0\linewidth]{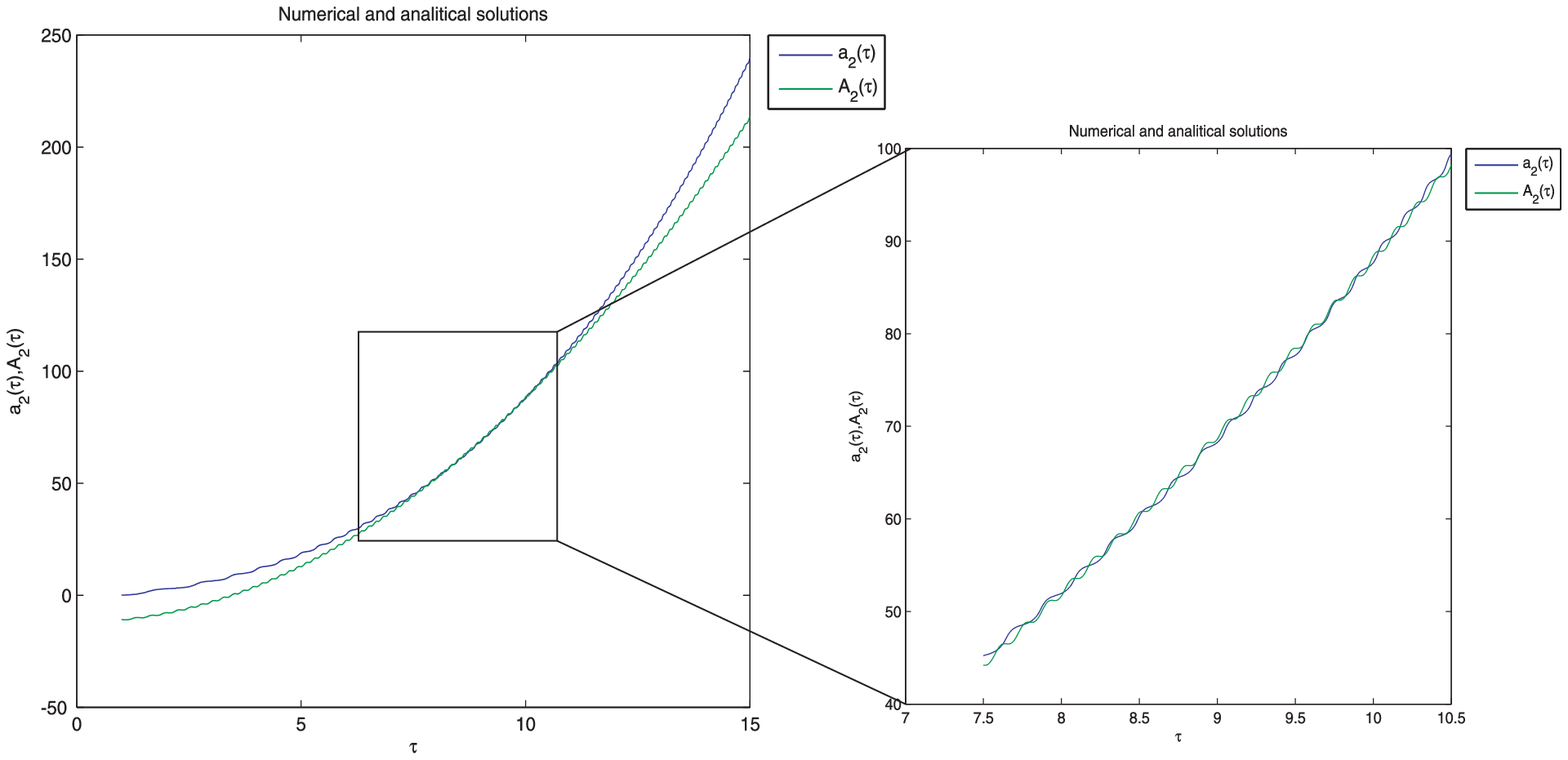} \\ a)}
\end{minipage}
\vfill
\begin{minipage}[h]{1.0\linewidth}
\center{\includegraphics[width=1.0\linewidth]{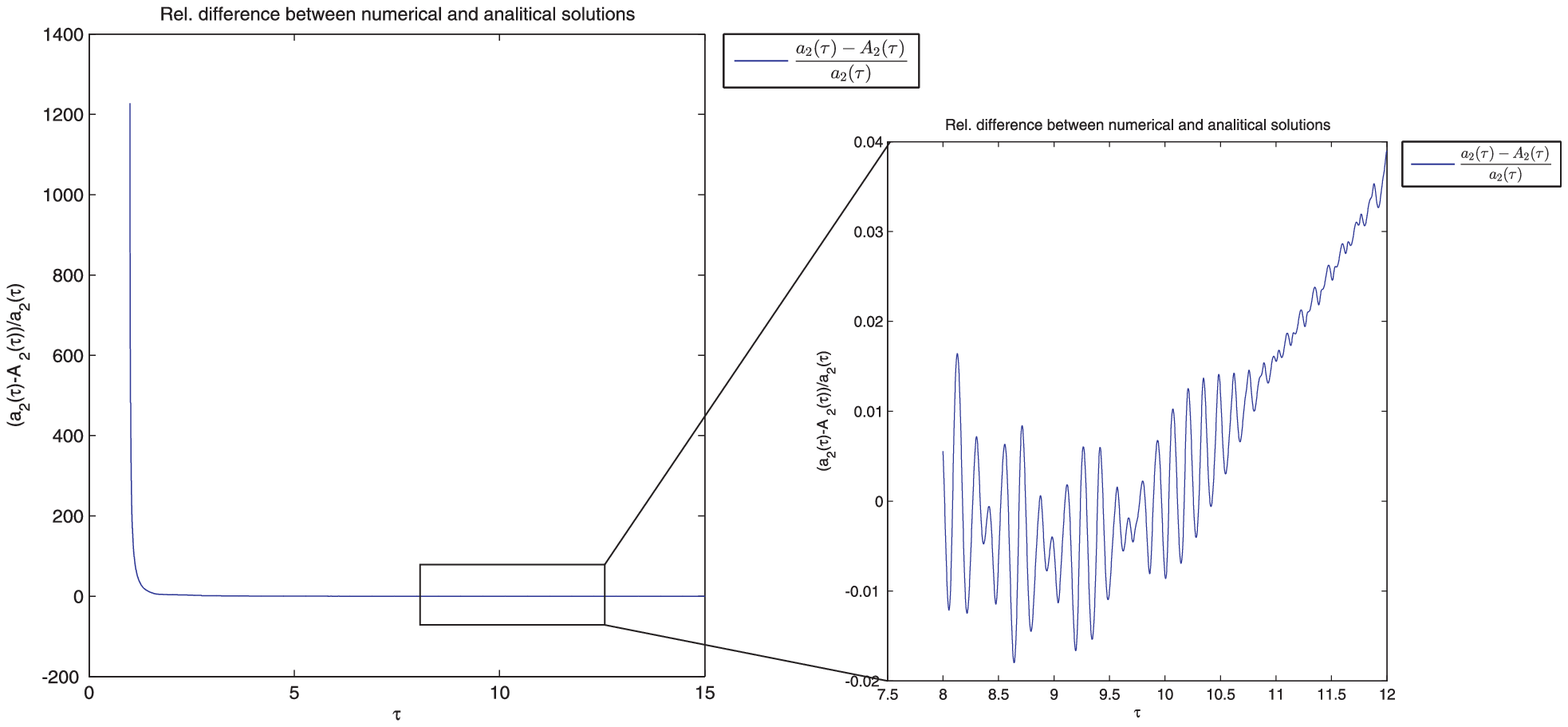} \\ b)}
\end{minipage}
\caption{\footnotesize a) Comparison of numerical ($a_2$) and analytical ($A_2$) solutions. b) Relative difference of numerical ($a_2$) and analytical ($A_2$) solutions.\;\;$\scriptsize A_1(\tau)=-\tau^2+0.7\tau J_{\frac{2}{3}}\left(10\tau^{\frac{3}{2}}\right)+0.7\tau Y_{\frac{2}{3}}\left(10\tau^{\frac{3}{2}}\right)+15.2$.}
\label{num-anlt-1}
\end{figure}

\begin{figure}[h]
\begin{minipage}[h]{1.0\linewidth}
\center{\includegraphics[width=1.0\linewidth]{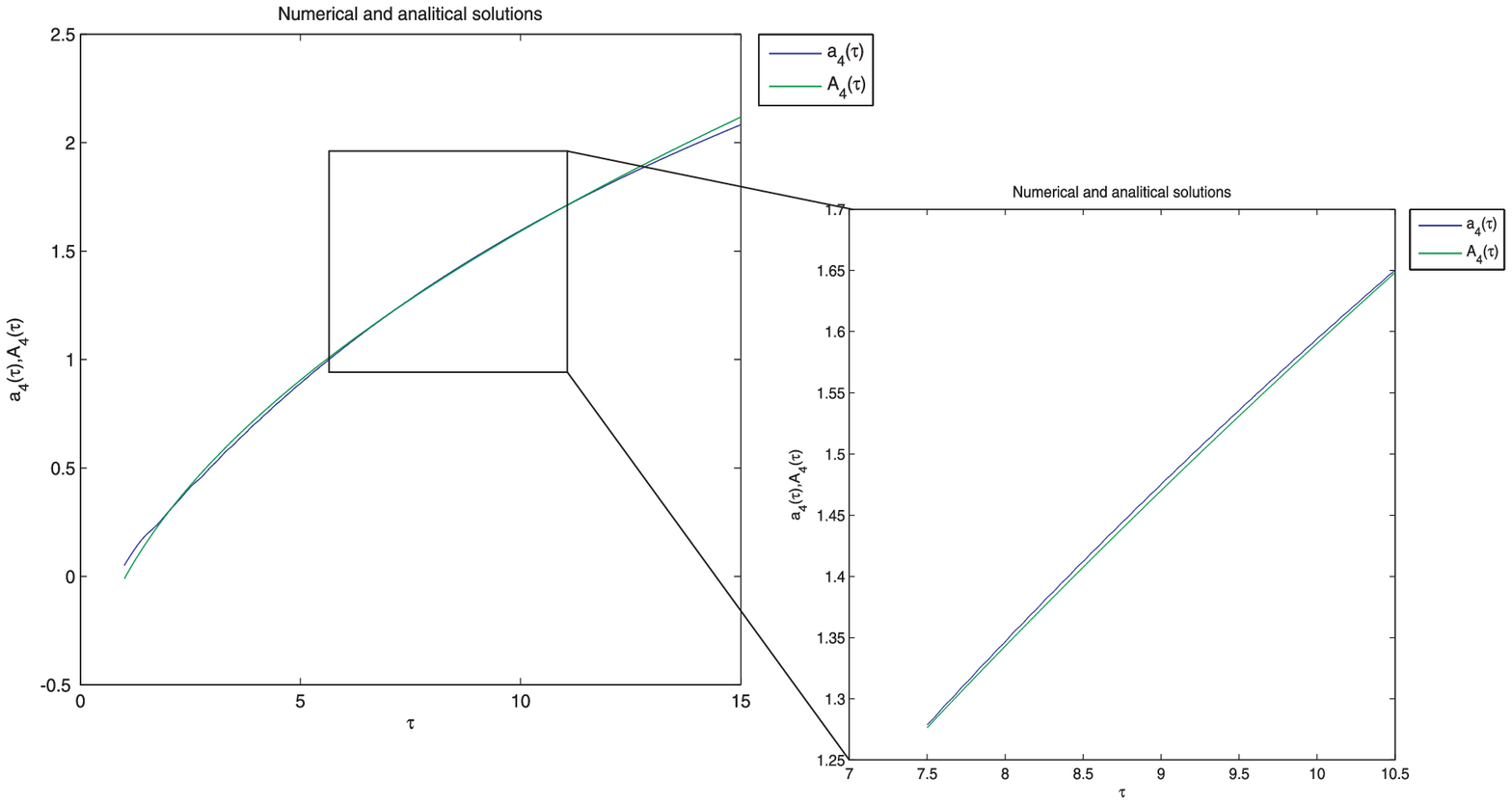} \\ a)}
\end{minipage}
\vfill
\begin{minipage}[h]{1.0\linewidth}
\center{\includegraphics[width=1.0\linewidth]{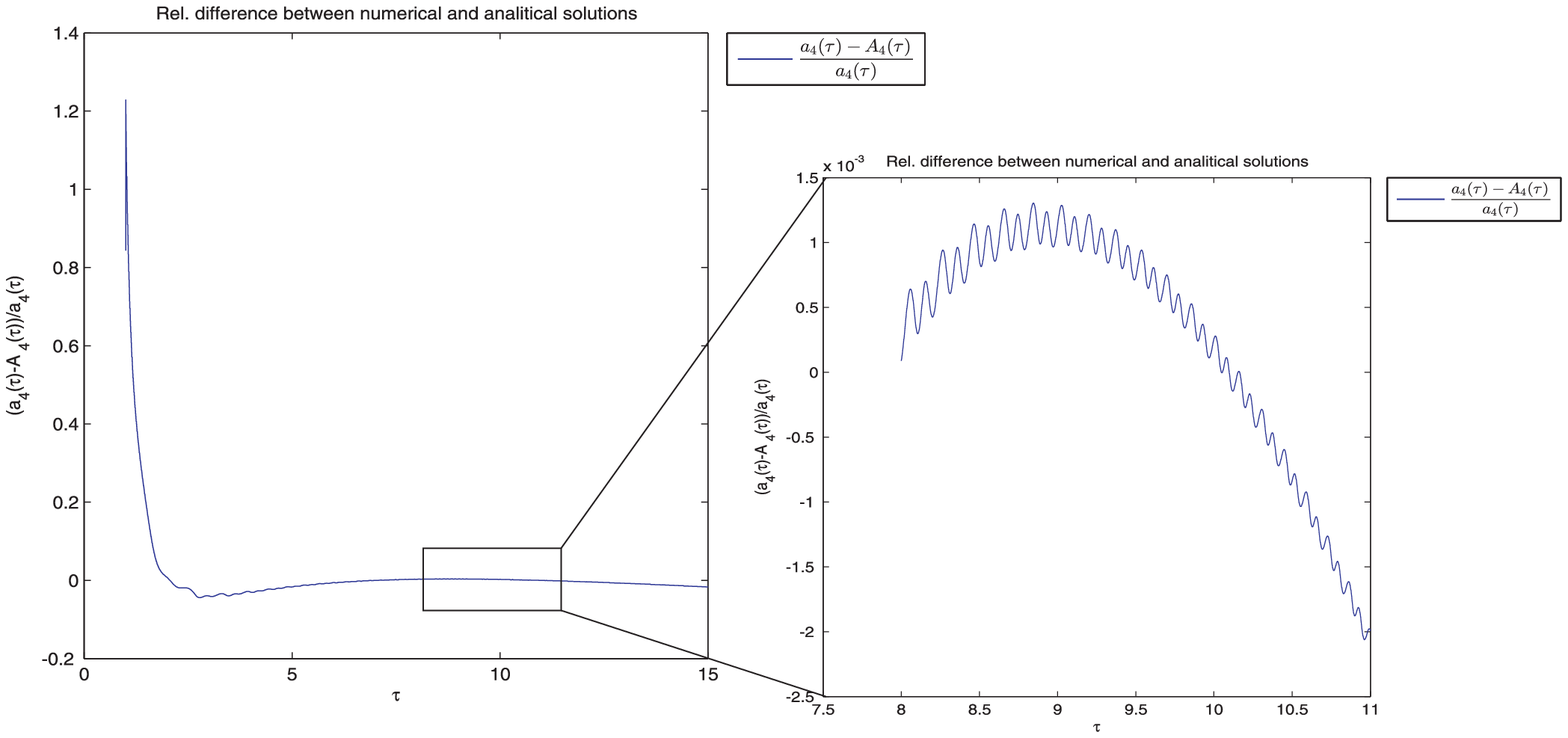} \\ b)}
\end{minipage}
\caption{\footnotesize a) Comparison of numerical ($a_4$) and analytical ($A_4$) solutions. b) Relative difference of numerical ($a_4$) and analytical ($A_4$) solutions.\;\;$A_4(\tau)=0.741\sqrt{\tau}-0.749$.}
\label{num-anlt-3}
\end{figure}

\section{Conclusions~\label{6}}

In this paper we provided an analysis of stability of known power-law solutions in Gauss-Bonnet anisotropic cosmology
in the presence of a homogeneous magnetic field in the case when metric is diagonal in the frame determined by magnetic field. The relative simplicity of equations of motion in this case allows us to find analytically a simple condition for the particular power-law solution to be stable. It appears that a set of
initial conditions for which the power-law solution can not be an attractor near a cosmological singularity has non-zero
measure (though does not coincide with the whole initial conditions space). The result has the same structure as in
the Einstein gravity (particular sums of Kasner indices should be restricted by some number), the only difference
is the exact value of this number.

The fate of trajectories near singularity from this unstable domain requires future investigations. In the present paper
we describe one particular class of trajectories found by numerical studies. The volume element experiences oscillations
in the presented regime, making it qualitatively different from  BKL oscillations.

We can expect that some other regimes which do not tend to Kasner-like power-law regime exist in the system
under investigation. Full description of them (the task which we leave for a future work) in this particular rather simple
model can be useful in studies of cosmological singularity in Lovelock gravity in the presence of spatial curvature.
The latter model usually lead to very cumbersome equations of motion, so understanding the nature of possible regimes
in the instability zone for the Kasner-like solution in the present model would help in finding similar solution in
more general and more technically complex situations of curved multidimensional geometries.

\section*{Acknowledgments}

The authors are grateful to Vladimir Ivashchuk for helpful discussion. The work is partially supported by the RFBR grant 11-02-00643.


\begin{thebibliography}{99}

\bibitem{BKL} V.A.Belinskii, I.M.Khalatnikov and E.M.Lifshitz, Adv.Phys. 19, 525 (1970).

\bibitem{LeBlanc} Victor G. LeBlanc, Class.Quantum.Grav. 14, 2281 (1997).

\bibitem{Kirillov} R. Benini, A.A. Kirillov, G. Montani, Class.Quant.Grav. 22, 1483-1491 (2005).

\bibitem{Rosen} G. Rosen, J.Math.Phys. 3, 313 (1962).

\bibitem{Deruelle1} N. Deruelle, Nucl.Phys. B 327, 253-266 (1989).

\bibitem{Deruelle2} N. Deruelle, L. Farina Busto, Phys.Rev. D 41, 3696 (1990).

\bibitem{Toporensky} A. Toporensky, P.Tretyakov, Grav.Cosmol. 13, 207-210 (2007); \href{http://arxiv.org/abs/0705.1346v3}{arXive: 0705.1346}.

\bibitem{Pavluchenko} S.A. Pavluchenko, Phys.Rev. D 80, 107501 (2009); \href{http://arxiv.org/abs/0906.0141v2}{arXiv: 0906.0141}.

\bibitem{Kirnos1} I. Kirnos, A. Makarenko, S. Pavluchenko and A. Toporensky, Gen.Rel.Grav. 42, 2633 (2010) \href{http://arxiv.org/abs/0906.0140v1}{arXive: 0906.0140}.

\bibitem{Kirnos2} I.Kirnos, S. Pavluchenko and A. Toporensky, Grav. Cosmol., 16, 274-282 (2010) \href{http://arxiv.org/abs/1002.4488v2}{arXive: 1002.4488}.

\bibitem{Ivashchuk1} V. Ivashchuk, Grav.Cosmol., 16, 118-125 (2010) \href{http://arxiv.org/abs/0910.3426v3}{arXive: 0910.3426}.

\bibitem{Ivashchuk2} V. Ivashchuk, Int.J.Geom.Meth.Mod.Phys., 7, 797-819 (2010) \href{http://arxiv.org/abs/0910.3426v3}{arXive: 0910.3426}.

\bibitem{Mitskievich1} N.V. Mitskievich, Rev.Mex.Fis., 49S2, 39-51 (2003) \href{http://arxiv.org/abs/gr-qc/0202032v2}{arXive: 0202032}.

\bibitem{Mitskievich2} N.V. Mitskievich, \emph{Relativistic Physics in Arbitrary Reference Frames}, Nova Science Publishers, 2006 \href{http://arxiv.org/abs/gr-qc/9606051v1}{arXive: 9606051}

\bibitem{Mitskievich3} N.V. Mitskievich, \emph{Electromagnetism and perfect fluids interplay in multidimensional spacetimes}, in Proc. MG11 (2006) \href{http://arxiv.org/abs/0707.3190v1}{arXive: 0707.3190}

\bibitem{Kasner} Kasner E., American Journal of Math., 43, 217-221 (1921).


\end{thebibliography}
\end{document}